\documentclass[10pt,journal]{IEEEtran}

\usepackage{colortbl}
\usepackage{balance}
\usepackage{xspace}
\usepackage[numbers,sort]{natbib}
\usepackage[colorlinks,allcolors=violet]{hyperref}
\usepackage{booktabs}
\usepackage{supertabular}
\usepackage{bbding} 
\usepackage{orcidlink}

\makeatletter
\@namedef{ver@lineno.sty}{9999/12/31}
\@namedef{opt@lineno.sty}{}
\makeatother
\usepackage[frozencache,cachedir=minted-output]{minted}

\usepackage{algorithm}
\usepackage{algpseudocode}
\usepackage{multicol, multirow}
\usepackage{amsthm}
\usepackage{amsmath}
\usepackage{amssymb}
\usepackage[font=footnotesize]{caption,subcaption}
\usepackage{tablefootnote}
\usepackage{tcolorbox}
\usepackage{xcolor}
\usepackage[normalem]{ulem}
\usepackage{scrextend}
\usepackage{mathtools}
\usepackage{proof}
\usepackage{setspace}

\newcommand{\ourtoolraw}{ExpressAPR}
\newcommand{\ourtool}{\ourtoolraw{}\xspace}
\newcommand{\accOverPlain}{137.1}
\newcommand{\accOverSota}{8.8}
\newcommand{\ted}{mutant deduplication\xspace}

\newcommand{\napp}{execution scheduling\xspace}
\newcommand{\Napp}{Execution scheduling\xspace}
\newcommand{\NApp}{Execution Scheduling\xspace}
\newcommand{\napper}{execution scheduler\xspace}
\newcommand{\NApper}{Execution Scheduler\xspace}
\newcommand{\tree}{state-transition tree\xspace}

\newcommand{\eins}{interception-based instrumentation\xspace}

\newcommand{\EIns}{Interception-based Instrumentation\xspace}

\newcommand{\code}[1]{{\small \texttt{#1}}}
\newcommand{\xyaadd}[1]{#1}

\newcommand{\xyadel}[1]{}

\newcommand{\figref}[1]{Figure~\ref{#1}}
\newcommand{\tabref}[1]{Table~\ref{#1}}
\newcommand{\secref}[1]{Section~\ref{#1}}

\newcommand{\algoref}[1]{Algorithm~\ref{#1}} 
\newcommand{\procref}[1]{Process~\ref{#1}}

\begin{document}

\title{Accelerating Patch Validation for Program Repair with Interception-Based \NApp}

%
%

\author{%
Yuan-An Xiao \orcidlink{0000-0002-5673-3831}, 
Chenyang Yang \orcidlink{0000-0001-5016-7296}, 
Bo Wang \orcidlink{0000-0001-7944-9182} and 
Yingfei Xiong \orcidlink{0000-0001-8991-747X} 
\thanks{
\IEEEcompsocthanksitem{}This work was supported by the National Key Research and Development Program of China under Grant
No. 2022YFB4501902, the National Natural Science Foundation of China under Grant No. 62202040 and 62161146003, and ZTE Industry-University-Institute Cooperation Funds
under Grant No.HC-CN-20210319008.
\IEEEcompsocthanksitem{}Yuan-An Xiao and Yingfei Xiong are with the Key Laboratory of High Confidence Software Technologies (Peking University), Ministry of Education; School of Computer Science, Peking University, Beijing 100871, China (e-mail: xiaoyuanan@pku.edu.cn; xiongyf@pku.edu.cn).
\IEEEcompsocthanksitem{}Chenyang Yang is with the School of Computer Science, Carnegie Mellon University (e-mail: cyang3@cs.cmu.edu). This work was finished mostly when Chenyang Yang was an undergraduate student at Peking University.
\IEEEcompsocthanksitem{}Bo Wang is with the School of Computer and Information Technology, Beijing Jiaotong University, Beijing 100044, China (e-mail: wangbo\_cs@bjtu.edu.cn).
}
}

%
%

\markboth{IEEE TRANSACTIONS ON SOFTWARE ENGINEERING}%
{Accelerating Patch Validation for Program Repair with Interception-Based \NApp}
%



\IEEEtitleabstractindextext{
\begin{abstract}
Long patch validation time is a limiting factor for automated program repair (APR). Though the duality between patch validation and mutation testing is recognized, so far there exists no study of systematically adapting mutation testing techniques to general-purpose patch validation. To address this gap, we investigate existing mutation testing techniques and identify five classes of acceleration techniques that are suitable for general-purpose patch validation. Among them, mutant schemata and \ted have not been adapted to general-purpose patch validation due to the arbitrary changes that third-party APR approaches may introduce. This presents two problems for adaption: 1) the difficulty of implementing the static equivalence analysis required by the state-of-the-art \ted approach; 2) the difficulty of capturing the changes of patches to the system state at runtime. 


To overcome these problems, we propose two novel approaches:
1) \napp, which detects the equivalence between patches online, avoiding the static equivalence analysis and its imprecision;
2) \eins, which intercepts the changes of patches to the system state, avoiding a full interpreter and its overhead.



Based on the contributions above, we implement \ourtool, a general-purpose patch validator for Java that integrates all recognized classes of techniques suitable for patch validation.
Our large-scale evaluation with four APR approaches shows that \ourtool accelerates patch validation by \accOverPlain{}x over plain validation or \accOverSota{}x over the state-of-the-art approach, making patch validation no longer the time bottleneck of APR. Patch validation time for a single bug can be reduced to within a few minutes on mainstream CPUs.
\end{abstract}

}

\maketitle

\IEEEdisplaynontitleabstractindextext

%
\IEEEpeerreviewmaketitle

\section{Introduction}

\IEEEPARstart{A}{utomated} program repair (APR) has attracted much attention in the recent decade.
Many approaches~\cite{le2016history,liu2019tbar,weimer2009automatically,mechtaev2018semantic,liu2013r2fix,kim2013automatic,wei2010automated,zhu2021syntaxguided}
have been proposed, and companies like Bloomberg~\cite{kirbas2021introduction}, Meta~\cite{marginean2019sapfix}, and Alibaba~\cite{zhang2020precfix} are already using APR tools to fix software bugs in nightly builds.

Though researchers have made significant progress in the effectiveness of APR, its efficiency has received relatively limited improvement. Efficiency decides the time needed for repairing a bug and is an important limiting factor for using APR in practice. The state-of-the-art APR approaches may still take hours to repair a bug, and recent program repair experiments still set the timeout to multiple hours~\cite{zhu2021syntaxguided,liu2019tbar,noller2022trust}.
The response time of current APR tools greatly exceeds users' patience, as reported by recent studies~\cite{noller2022trust,liang2021interactive}, and significantly limits the application scenario of APR approaches.

The execution time of APR is dominated by patch validation~\cite{le2013current,ghanbari2020prf}.
Most APR approaches are test-based and follow the generate-and-validate pattern~\cite{le2011genprog}: they first generate a bunch of patches, and then validate each patch using the test suite.
Executing the test suite can take minutes, and hundreds of patches can be generated for one fault~\cite{liu2020efficiency}. To accelerate
APR tools, we need to reduce the patch validation time.

Patch validation has been recognized as a dual of mutation testing~\cite{weimer2013leveraging}. Mutation testing applies a set of mutation operators to the program to produce mutants, and then executes the test suite on each mutant to calculate the mutation score.
Mutation testing is a dual of patch validation as both need to obtain the test result for each mutant/patch.
Given many techniques have been proposed to accelerate mutation testing, they could potentially be adapted to accelerate patch validation.

Under this duality, multiple existing attempts to accelerate patch validation \xyaadd{can be seen as} adapting the corresponding mutation testing technique~\cite{weimer2013leveraging,chen2017contract,mechtaev2018test,hua2018practical,chen2021fast,ghanbari2020prf,guo2019speedup,mehne2018accelerating,qi2013efficient,ghanbari2019practical,wong2021varfix,fast2010designing,pacheco2007feedback}. However, all
of them adapt only one or a few techniques. It remains unclear which techniques can be adapted and what is the combined effect of these techniques on patch validation. Furthermore, many attempts are special-purpose, depending on a specific APR approach and not generalizing to different ones.

To fill this gap, we first systematically investigate the existing acceleration techniques for mutation testing and analyze their suitability for patch validation.
We identify five classes of suitable techniques, namely,
mutant schemata~\cite{untch1993mutation},
\ted~\cite{just2011major},
test virtualization~\cite{bell2014unit},
test prioritization~\cite{zhang2013faster}, 
and parallelization.
The remaining techniques are either covered by these five classes or are unsuitable for the patch validation of current APR approaches.
Among them, the latter three classes are already used by recent general-purpose patch validators~\cite{chen2021fast,ghanbari2020prf}, but mutant schemata and \ted have never been adapted to a general-purpose patch validator, as far as we are aware. Mutant schemata weave all mutants into one program and avoid redundant compilation of different mutants; \ted detects mutants that are equivalent to each other, and executes only one mutant among all equivalent mutants. 

Migrating these two techniques to patch validation is not easy. In mutant testing, mutants are produced from pre-defined mutation operators, and their impact on the system is under control. 
However, in patch validation, the patches are produced from third-party APR approaches that arbitrarily change the program.
This imposes two problems:

\textbf{The first problem is that the state-of-the-art \ted approach~\cite{just2014efficient} requires a static analysis} to determine the equivalence of the rest of the test executions, when two patches deviate from the original program. The precision of this analysis is directly related to the effectiveness of the technique. It is difficult to implement a precise and scalable static equivalence analysis given the possible changes that the patches may cause to the system.

To address this problem, we propose a novel approach: \textbf{\napp}. \Napp does not require offline static equivalence analysis, and subsumes mutant schemata and \ted.
The basic idea is to detect the equivalence between patches online during patch execution. Doing so removes the need for static equivalence analysis, thereby preventing the imprecision of static analysis from affecting the system’s effectiveness. We also introduce a novel data structure, the \emph{\tree}, to record the dynamic analysis results and reuse them across the execution of different patches.
Furthermore, to perform the detection, we need to weave all patches into a single program, subsuming mutant schemata. This approach is compatible with test prioritization, test virtualization, and parallelization, allowing us to integrate all suitable techniques into one system.

\textbf{The second problem is that implementing mutant deduplication or execution scheduling requires a component to capture how a patch changes the system state}
to analyze equivalence and replay the change later.
Existing mutant deduplication approaches
usually implement an interpreter to execute the changed code.
However, unlike mutation testing where only a selected set of operators may be changed and interpreted, a patch may arbitrarily change a statement that may involve method calls or even system calls. As a result, the interpreter must support all features in the host programming language and interact with the original runtime. This not only requires huge implementation effort but is also technically difficult as the original runtime may be commercial and prohibit modification. Furthermore, interpreting a large chunk of code inevitably incurs a significant overhead, which may nullify the effectiveness of the technique. 

\xyaadd{To support the implementation of execution scheduling, }
we propose a novel approach for \textbf{\eins}. It executes a patch without an interpreter by instrumenting code before and after the patch to record the possibly changed states and revert the changes. In this way, the patch is executed in the original runtime, avoiding the implementation cost and the runtime overhead of an interpreter. 

{
However, realizing \eins is not easy. We highlight our approach in three aspects:
1) \textbf{A patch may bring control flow changes} via statements like \code{break} and \code{throw}. Capturing such changes is difficult because the instrumented statement may be skipped. We propose a design process based on the operational semantics of the programming language to reliably capture and replay the control flow change of a patch.
2) \textbf{The scope of data changes} for a patch is large, because a patch may call other methods and arbitrarily modify the memory at any location. The cost of recording all possible changes may therefore cancel the benefit of \ted. So, we analyze the possibly changed locations in a preparation step and apply equivalence detection only when the change scope is small enough.
3) \textbf{A patch may not compile}, and weaving all patches together may make the whole program uncompilable. We cannot simply detect an uncompilable patch by compiling it, because the benefit of mutant schemata will be lost. Instead, we propose the concept of \emph{isolation unit} where compilation errors within one unit would not affect other units. We separate each patch into an isolation unit and identify all uncompilable patches at once.
}

\textbf{Based on the contributions above, we implement \ourtool~\cite{expressapr_replication},} a general-purpose patch validator for Java, which integrates all suitable techniques for patch validation. \ourtool provides support for Defects4J and Maven, and allows configuration for other Java projects. We then conduct a large-scale empirical evaluation that consumes over 17 months of CPU time to understand its performance on mainstream APR approaches. The evaluation leads to multiple findings:

\begin{itemize}
    \item \ourtool achieves an acceleration of \accOverPlain{}x over plain validation and outperforms the state-of-the-art approach by \accOverSota{}x. Patch validation is now faster than patch generation for the first time. The time to repair a bug can be reduced to a few minutes, meeting the expectation of most developers~\cite{noller2022trust}.
    \item All adapted techniques are effective for patch validation, each contributing to a significant acceleration.
    \item \xyaadd{While \ourtool only works for patches under certain assumptions (Section~\ref{section:limitation}), it supports over 97\% of patches in our experiment, and it brings correct validation results for over 99.9\% of patches. Therefore, \ourtool has a negligible impact on APR effectiveness.}    
\end{itemize}

In summary, the contributions of this paper are twofold:

\textbf{Technically,} we propose two novel approaches to realize mutation schemata and \ted, which have not been adapted to general-purpose patch validation before. The \napp approach (\secref{section:approach}) subsumes mutant schemata and \ted, avoiding the need for a static equivalence analysis that can be imprecise. The \eins approach (\secref{section:approach_patch_virtualization}) provides the capture-and-replay component required by \napp, avoiding the need for a full-featured interpreter with runtime overhead.

\textbf{Empirically,} we investigate existing techniques for accelerating mutation testing, and identify five classes of techniques suitable for patch validation (\secref{section:survey}). We systematically integrate the complete set of techniques (\secref{section:approach_impl}) and conduct a large-scale experiment evaluating their performance (\secref{section:exp_setup}), leading to novel findings that can serve as a solid base for future research (\secref{section:exp_result}).


\section{Motivation}
\label{section:motivation}

Efficiency is a critical factor of APR. How long an APR tool takes to repair a bug determines the scenarios where it can be used. \citet{noller2022trust} recently found that half of programmers cannot wait longer than 30 minutes for an APR tool to produce a result. They further investigated how existing APR tools on the C programming language performed within one hour, revealing that even the best tool could correctly repair only two bugs in the ManyBugs benchmark. Such performance is significantly lower than the results reported in their original papers with a timeout of 10-24 hours. 

Due to this unsatisfactory efficiency, it is commonly believed that the application of APR can only be applied to offline scenarios where users do not wait for real-time feedback (e.g., repairing bugs discovered in automatic nightly builds~\cite{liang2021interactive,noller2022trust}). If we can reduce the time of program repair to a few minutes, APR tools can assist users with real-time feedback, unlocking many more application scenarios (e.g., being integrated into IDEs and code editors).

To improve the efficiency of APR tools, we need to understand how current APR tools spend their time. This paper focuses on generate-and-validate APR, which is the subject of mainstream studies. These approaches have two main phases: in the \emph{patch generation} phase, they loop through each suspicious location and generate a set of candidate patches at each location; in the \emph{patch validation} phase, they execute the test suite to check the plausibility of each patch. The patch validation phase can be further divided into \emph{patch compilation} and the \emph{test execution} for compiled languages.

To understand how much time existing APR tools spend on different phases, we conducted a pilot study on the state-of-the-art tools. For patch generation, we selected Recoder~\cite{zhu2021syntaxguided}, an effective APR approach based on deep learning. For patch validation, we selected UniAPR~\cite{chen2021fast}, a state-of-the-art patch validator. We took all bugs in the Defects4J 1.2 benchmarks that Recoder successfully repairs~\cite{zhu2021syntaxguided}. For each bug, we first executed the patch generation phase of Recoder with default parameters, and then used UniAPR to validate them. Since Defects4J provides a default validator (\code{defects4j compile \&\& defects4j test}), which is used in many existing APR tools, we also executed the Defects4J validator for comparison. 

\begin{table}[t]
    \centering\footnotesize
    \caption{
        Average time usage per bug when evaluating Recoder}
    \label{tab:motivation_timing_each_step}
    \begin{tabular}{lll@{\hspace{2em}}r}
        \toprule
        Phase & Step & Technique              & Time \\
        \midrule
        \multicolumn{2}{l}{Patch Generation}& Recoder  & 0:18:39 \\ 
        \midrule
        \multirow{4}{*}{Patch Validation} & \multirow{2}{*}{Patch Compilation} & Defects4J & 2:36:04 \\ 
                                     && UniAPR     & 0:25:34 \\ 
                                     \cline{2-4}\addlinespace[0.8mm]
        &\multirow{2}{*}{Test Execution} & Defects4J    & 19:53:48 \\ 
                                   && UniAPR      & 1:16:37 \\ 
        \bottomrule
    \end{tabular}
\end{table}

We recorded the average time used in each step, as shown in \tabref{tab:motivation_timing_each_step}. We can see that the patch validation phase is the time bottleneck, accounting for most of the time (84.57\% for UniAPR and 98.64\% for Defects4J) and making the total repair time much longer than the 30-minute expectation. Therefore, to improve the efficiency of APR tools, we need to reduce the patch validation time. Furthermore, both steps in patch validation cost a significant amount of time even with the best technique. Therefore, to reduce the patch validation time, we need to reduce both the patch compilation time and the test execution time. 

\section{Investigating Mutation Testing Acceleration}
\label{section:survey}

We review existing mutation testing literature to systematically find acceleration techniques applicable to patch validation. We first collect mutation testing acceleration techniques by reading through the existing surveys~\cite{jia2010analysis,papadakis2019mutation,pizzoleto2019systematic}, searching for publications matching the term ``mutation testing'' after 2019, after which the papers were not covered by the surveys, and studying the web page collecting existing mutation testing tools~\cite{Pitest_java_mutation_testing_systems}. We then analyze the suitability of each technique for accelerating patch validation.

Our analysis identifies a set of techniques that are suitable to be adapted to patch validation. The remaining techniques are either unsuitable for validating the patch generated by current APR tools, or covered by the identified set of techniques. In this section, we present the classification result and briefly introduce each technique to make the paper self-contained.

\subsection{Techniques Suitable for Patch Validation}
\label{section:survey_used}

\subsubsection{Mutant Schemata}
\label{section:survey_mutation_schemata}
In mutation testing, each generated mutant needs to be compiled. Different mutants share most of their code, which is repetitively compiled. 
Mutant schemata \cite{untch1993mutation} is a common technique to avoid repetitive compilation.
Multiple mutants are encoded in a \textit{meta-program} and then dynamically selected during runtime, so the shared code is compiled only once, reducing the redundant compilation.

\figref{fig:motivation_example_mutation_schemata} illustrates an example where the AOR
(replace arithmetic operator) mutation operator is applied to the expression \code{a+b}, as shown in (a).
AOR generates multiple mutants by replacing the plus sign with other operators, 
as shown in (b). Normally each mutated program is independently compiled, so other parts of the code are compiled multiple times. 
But with mutant schemata, all mutants at this expression are grouped into a \textit{meta-function} that dynamically selects a mutant based on a runtime flag, as shown in (c).
In this way, all mutants are encoded in one program, and other parts of the code are compiled only once.
Mutant schemata can be generated 
by source code (AST) \cite{just2011major} or byte-code \cite{ma2005mujava} transformation.

\begin{figure}[t]
    \centering
    
    \begin{subfigure}[b]{0.2\textwidth}
        \begin{minted}[fontsize=\scriptsize]{java}
        c = a + b;
        \end{minted}
        \vskip -.75em
        \caption{Program to be mutated}
        
        \vskip .5em
        
        \begin{minted}[fontsize=\scriptsize]{java}
      // mutant 1:
        c = a - b;
      // mutant 2:
        c = a * b;
      // mutant 3:
        c = a / b;
        \end{minted}
        \vskip -.75em
        \caption{Generated mutants}
    \end{subfigure}
    \begin{subfigure}[b]{0.25\textwidth}
        \begin{minted}[fontsize=\scriptsize]{java}
        
  c = metaFunc(a, b);
  int metaFunc(int x, int y) {
    switch(Env.getMutId()) {
      case 1: return x - y;
      case 2: return x * y;
      case 3: return x / y;
    }
  }
        \end{minted}
        \vskip -.5em
        \caption{The mutant schemata}

    \end{subfigure}
    
    \caption{
        Mutants and mutant schemata created by AOR
    }
    \label{fig:motivation_example_mutation_schemata}
\end{figure}

\subsubsection{Mutant Deduplication}
\label{section:survey_dyn_dedup}

Since a mutant is created by mutating only one or a few statements in the original program, a mutant may be equivalent to the original program or another mutant. In a group of equivalent mutants, only one of them needs to be executed, saving the test execution time. Two types of equivalencies have been considered in existing work. The first one is \emph{full equivalence}, where the two mutants will produce the same test result on any test. For example, the statement \code{x+=2;} and \code{x+=1+1;} are fully equivalent. Given a group of fully equivalent mutants, only one needs to be executed on all tests. The second one is \emph{test-equivalence}, where the two mutants will produce the same test result on a specific test. Test-equivalence is much more common than full equivalence. For example, the statement \code{x+=2;} and \code{x*=2;} are not fully equivalent but are test-equivalent if \code{x} equals 2 before invoking the mutated statement in a test. Furthermore, any two mutants whose mutated statements are not executed by a test are equivalent with respect to the test. Given a group of test-equivalent mutants with respect to test $t$, only one needs to be executed on $t$. 

To deduplicate mutants, existing approaches employ an offline procedure to detect equivalent mutants before the test execution, and then select one mutant among each equivalence class to be tested.
\citet{baldwin1979heuristics} and \citet{papadakis2015trivial} utilize compiler optimization to detect fully equivalent mutants, while \citet{pan1994using} uses constraint solvers to identify the fully equivalent mutants. 

Since test-equivalence is more common than full equivalence, 
the Major mutation framework~\cite{just2014efficient} identifies test-equivalent mutants by a pre-pass executing the original program and interpreting the mutated expressions in the mutants along the execution. Since the pre-pass is performed on the original program, Major can successfully detect test-equivalence between the original programs and the mutants, but for test-equivalence between mutants, Major requires a static analysis to determine the equivalence for the rest of the executions after they deviate from the original program. More discussion can be found in the next section.


\subsubsection{Test Virtualization}
\label{section:survey_vmvm}

Before executing a test for the mutants, the test suite must be initialized for each mutant. 
The repetitive initialization is costly in languages based on a virtual machine (VM), such as Java, as booting the VM takes non-trivial time. 
Test virtualization approaches such as VMVM~\cite{bell2014unit} reduce this cost by reusing the previous VM instance for the next mutant execution. 
The global 
variables changed by the previous test execution are identified and reset before the next round of test execution by instrumentation.

\subsubsection{Test Case Prioritization}
\label{section:survey_test_prio}

The idea of test case prioritization is that some test cases run faster or are more likely to fail,
so that if these test cases run before other test cases, the mutant will be killed earlier when they fail~\cite{zhang2013faster,yoo2012regression}.
Different heuristic policies can be used to determine what tests get prioritized. In particular, in \textit{regression testing}, of which APR patch validation can be seen as an instance, a typical heuristic is to prioritize the previously failed test cases. Another heuristic is to prioritize the test cases in the same package of the modified code, which are most likely to cover the modified code.

\subsubsection{Parallelization}
Many mutation testing tools~\cite{Pitest_java_mutation_testing_systems,schuler2009javalanche} seek parallel test execution on multicore processors. Because tests and mutants are independent of each other, parallelization can be trivially applied by dividing the work into multiple pieces to be consumed by a process pool, e.g., each piece dealing with a small group of mutants or tests. Some early-stage mutation testing approaches also utilize hardware parallelization mechanisms, such as SIMD~\cite{krauser1991high} and MIMD~\cite{offutt1992mutation}.

\subsection{Techniques Unsuitable for Current APR}
A class of acceleration techniques uses an optimized execution engine to execute all mutants at the same time, reducing possible redundant execution when executing each mutant separately. We classify these techniques as unsuitable for general-purpose patch validation for current APR approaches, because they either require a specific platform feature, or incur too much overhead such that the benefit gained is difficult to surpass the overhead:

\textit{Fork-based mutation analysis} tries to accelerate by sharing the same execution among mutants, relying on the fork mechanism of the POSIX systems.
Spilt-stream execution (SSE) begins with one process representing all mutants, and then forks into multiple subprocesses when a mutated location in the program is reached, one subprocess representing one mutated statement~\cite{tokumoto2016muvm,gopinath2016topsy}. AccMut~\cite{wang2017faster} and WinMut~\cite{wang2021faster} are two enhanced versions of SSE. When a mutated location is reached, their engines analyze all mutants for their changes to the system state, and cluster the mutants based on their changes: the mutants whose changes lead to the same system state are in the same cluster. Finally, the engines fork only one process for each cluster, reducing the number of processes compared with SSE. However, all of the fork-based approaches require the fork mechanism, and are not suitable to be used in a general-purpose patch validation approach as we need to support different operating systems and programming languages that may not support the fork mechanism, such as Windows and Java.

\textit{Variational execution}~\cite{wong2018faster} employs a special execution engine allowing a variable to store a conditional value, which is a special data structure that records possible values of the variable in all mutants. Operations in the program, such as plus or minus, directly manipulate conditional values, and thus executing the program with conditional values produces the test result of all mutants. However, executing the statements over conditional values incurs significant overhead, and thus variational execution is effective only when the number of mutants is huge (e.g., exponential). As reported in an existing study~\cite{wong2018testing}, variational execution could slow down the execution when there are only tens of or hundreds of mutants, a scenario commonly encountered in patch validation. 

Furthermore, all techniques discussed above are \textit{lossless} acceleration techniques. There are also \textit{lossy} acceleration techniques~\cite{jia2010analysis,mathur1994empirical,zhang2018predictive,howden1982weak,offutt1994empirical,wong1995reducing,offutt1993experimental,ji2009novel}, which approximate the test results on the mutants to produce the mutation score, which is defined as the ratio of mutants that cause a test failure. However, on patch validation, we need to get the accurate test result on each patch to identify a correct patch, rather than calculating a ratio, and thus the lossy techniques do not apply to the patch validation.

\subsection{Techniques That Are Subsumed}
\label{section:survey_not_used}

\textit{Test case selection} is an optimization used in
several mutation testing tools \cite{Pitest_java_mutation_testing_systems,gligoric2015practical,harrold2001regression} that skips test cases that
do not execute the mutated code. It is subsumed by \ted because
if the test does not execute the mutated code, all mutants should be in a test-equivalence class.

\section{\NApp}
\label{section:approach}

\newtheorem*{theorem_soundness}{Theorem (Soundness)}
\newtheorem*{theorem_efficiency}{Theorem (Efficiency)}
\algrenewcommand\algorithmicindent{1em}

\newcommand{\step}[4]{\ensuremath{{#1}  \xrightarrow{{#2}}\langle {#3}, {#4} \rangle}}
\newcommand{\eval}[3]{\ensuremath{{#1}  \xrightarrow{{#2}}{#3}}}

\subsection{Overview}
\label{section:approach_overview}

Among all the five classes of suitable acceleration techniques, mutant deduplication and mutant schemata have not been adapted to general-purpose patch validation due to the differences between mutation testing and patch validation. In this section, we propose the \napp approach, which is suitable for general-purpose patch validation and subsumes both acceleration techniques.

We first demonstrate the possible redundancies during patch validation with an example as shown in the code snippet below. In this code snippet are five patches, $P_1$ to $P_5$, each modifying either of the two statements in the function \code{f()} ($S_1$ for the former two patches, $S_2$ for the latter three patches). The test function \code{test()} determines the correctness of \code{f()} by asserting its behavior.

\begin{minted}[fontsize=\scriptsize]{java}

    int i=2, j=1;
    void f() {
        i+=2; // S1
     // P1: "i*=2;" P2: "i=j+3;"
        j+=2; // S2
     // P3: "j*=2;" P4: "j=i-2;" P5: "j=2;"
    }
    void test() {
        f(); assert(i==4 && j==2);
        f(); assert(i==6 && j==4);
    }

\end{minted}

\begin{figure}[t]
    \centering
    \includegraphics[width=\columnwidth]{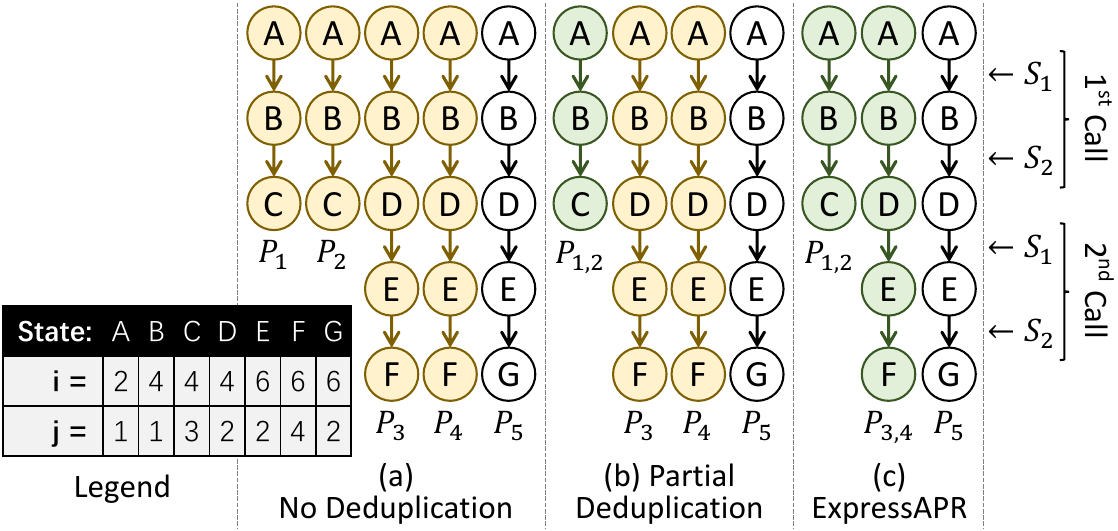}
    \caption{State transitions with different patch validation approaches}
    \label{fig:approach_test_exec}
\end{figure}

\figref{fig:approach_test_exec}(a) illustrates the redundant test executions in this example, where each vertical sequence shows the execution of the test on one patch, and each circle indicates the system state at a specific location. $P_1$ and $P_2$ are test-equivalent to the original program, because all of them effectively set \code{i=4} and \code{j=3} at the first call to \code{f()} and thus the first assertion in \code{test()} fails (State C). $P_3$ and $P_4$ are test-equivalent to each other, setting \code{i=4,j=2} at the first call to \code{f()} (State D), and \code{i=6,j=4} at the second call (State F), passing both assertions. $P_5$ is not test-euivalent to any other variant: though $P_5$ sets \code{i=4,j=2} at the first call to \code{f()}, temporarily leading to the same state D as $P_3$ and $P_4$, their states deviate during the second call where $P_5$ sets \code{j=2} instead of \code{4} (State G). Therefore, if we take the plain patch validation approach, where the test iteratively executes over $P_1$ to $P_5$, test-equivalent patches always follow the same path of state transition, leading to redundancies as highlighted in yellow.

As mentioned, the state-of-the-art \ted approach, Major, detects test-equivalence by a pre-pass on the original program, and executes the test on only one patch from each equivalence class. In this example, Major will instrument the original program and execute the instrumented program as a pre-pass. When a location modified by patches is reached, e.g., $S1$, the instrumented code will invoke an interpreter to interpret all the patches based on the system state of the original program, record and compare their changes to the system state. If a patch always makes the same changes to the system state as the original program, it is test-equivalent to the original program, and thus does not need to test. In this example, it detects that $P_1$ and $P_2$ are test-equivalent to the original program. However, it is hard to identify whether $P_3$, $P_4$, and $P_5$ are test-equivalent to each other. Although they make the same change to the system state in their first invocation (state B), their behavior in the second invocation (state E) is unknown, because state E deviates from the pre-pass which follows the original program ($A \to B \to C$). Therefore, Major requires a static analysis to determine the equivalence between the patches whose execution deviates from the original program. How to implement such a static analysis is not discussed in the original publication, and implementing a precise and scalable static analysis of test-equivalence between patches is difficult. In this case, if we cannot statically determine the test-equivalence between $P_3$ and $P_4$, we have to execute the tests on both of them, leading to \figref{fig:approach_test_exec}(b). 

%

\begin{figure}[t]
    \centering
    \includegraphics[width=.95\columnwidth]{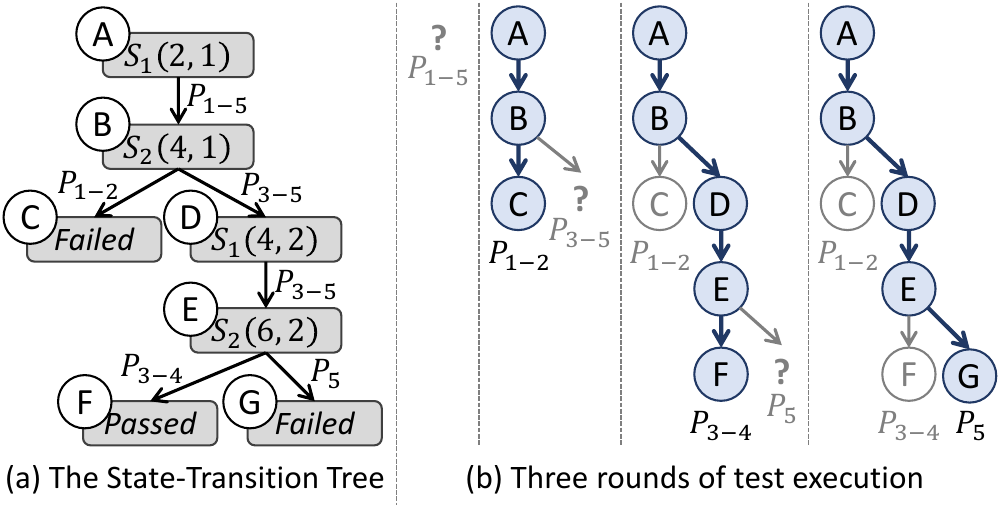}
    \caption{Iteratively building the state-transition tree in \ourtool}
    \label{fig:approach_state_transition_tree}
\end{figure}

Our \napp approach overcomes this limitation by embedding the detection process into the test execution process instead of in a pre-pass. To achieve that, we record the runtime behavior of patches as a \textit{state-transition tree}, as shown in \figref{fig:approach_state_transition_tree}(a). Each node in the tree indicates an \textit{interesting} program state of the test, where either the next statement to execute is changed by some patches, or the test completes. In the figure, a node annotated with $S_n (v_i, v_j)$ means that the statement $S_n$ will be executed with variables \code{i} and \code{j} set to $v_i$ and $v_j$, and a node annotated with ``Failed'' or ``Passed'' means the result of a completed test. Each edge $U \xrightarrow{P} V$ indicates the transition of interesting states: when executing the variant $P$ against the state $U$, the next interesting state will be $V$. For each patch, its execution corresponds to a path from the root node to a leaf node. 

The state-transition tree is built iteratively through multiple rounds of test execution. The current analysis result is recorded on the state-transition tree, so that in the next round we can test another patch that is not test-equivalent to any existing patches. As for the above example, $P_1$ to $P_5$ are validated with three rounds of test execution, as shown in \figref{fig:approach_state_transition_tree}(b). {In the first round, we randomly select a patch $P_1$, and throughout its execution, we record state changes of other patches onto the state-transition tree. We observe that $P_3$ to $P_5$ deviate from $P_1$ (with the state-transition path $A \to B \to C$) at $B$, as marked by the $?$ symbol. Therefore, in the second round, we select a patch among $P_3$ to $P_5$ to explore the deviated $A \to B \to \: ?$ path. This process repeats until all paths are explored.
In this way, we can achieve ideal mutant deduplication for this example, as illustrated in \figref{fig:approach_test_exec}(c), without the need for heavy static analysis in a pre-pass.}


\subsection{Problem Definition}
\label{section:approach_test_def}

Before introducing the approach, we define a set of concepts related to patch validation. These definitions abstract away details in different programming languages, such that our algorithm can apply to a wide range of programs that follow this definition.

We view the procedure of executing a test case as stepping through a state machine, as shown as \algoref{alg:exec_single}. The codebase consists of a set of locations and a mapping ($Cb . \text{Stmt}$) from locations to statements, which represents all source code in the project including tests. When a test begins, the location of the initial state $S$ points to the entry point statement of the test case ($Cb . \text{TestEntry}$). Then the statement gets executed, modifying $S$ that includes the current location ($S. \text{Loc}$). This process repeats until the state becomes a termination state that represents a failed or passed test result. Please note this algorithm is conceptual and does not imply an interpreter. In compiler-based language, the algorithm is implemented by the hardware architecture and the language runtime. 

\begin{algorithm}
    \caption{Test execution}
    \label{alg:exec_single}
    
    \begin{algorithmic}[1]
        \Statex \textbf{Input:} The codebase $Cb$
        \State $S \gets \{\text{Loc: } Cb . \text{TestEntry}\}$
        \While{$S$ is not a termination state}
            \State $S \gets \textsc{Execute}(S, Cb . \text{Stmt} [ S . \text{Loc} ])$
        \EndWhile
        \State $\textsc{ReportResult}(S)$
    \end{algorithmic}
\end{algorithm}

A patch is defined as a modification to the codebase, which is a modified mapping that replaces one or a few statements in the codebase. Given a set of patches $Ps$, a plain patch validation procedure enumerates through the patch set to execute the test against each modified mapping of statements, as shown as \algoref{alg:exec_plain}. If the codebase contains multiple test cases, this procedure is repeated for each test against patches that survive all previous tests.

\begin{algorithm}
    \caption{Plain patch validation}
    \label{alg:exec_plain}
    
    \begin{algorithmic}[1]
        \Statex \textbf{Input:} The codebase $Cb$, the patch set $Ps$
        \For{$P \in Ps$}
            \State $S \gets \{\text{Loc: } Cb . \text{TestEntry}\}$
            \While{$S$ is not a termination state}
                \State $S \gets \textsc{Execute}(S, P [ S . \text{Loc} ])$
            \EndWhile
            \State $\textsc{ReportResult}(P,\: S)$
        \EndFor
    \end{algorithmic}
\end{algorithm}

\subsection{The \NApper}

In this subsection, we explain how \napp works. Same as in the previous subsection, we shall only give an algorithm describing the process. How to implement it using instrumentation will be explained in \secref{section:approach_patch_virtualization}.

Our approach requires a component to capture the state changes of the patches for analysis. 
Concretely, we require that there exist a \textsc{Capture} procedure and a \textsc{Replay} procedure, so that $C \gets \textsc{Capture}(S, T)$ extracts the state change of statement $T$ over the state $S$ into $C$, and $S' \gets \textsc{Replay}(S, C)$ actually applies the state change $C$ to the state $S$. In this way, we can check whether the state change of two patches is the same by comparing the value of $C$. We require that these two procedures are accurate, such that $\textsc{Replay}(S, \textsc{Capture}(S, T)) = \textsc{Execute}(S, T)$ for any state $S$ and any statement $T$. 
Capturing and replaying changes will be discussed in \secref{section:approach_patch_virtualization}, and for now, we keep \textsc{Capture} and \textsc{Replay} as abstract procedures in this section.

\begin{algorithm}
    \caption{Patch validation with \napp}
    \label{alg:exec_scheduler}
    
    \begin{algorithmic}[1]
        \Statex \textbf{Input:} The codebase $Cb$, the patch set $Ps$
        \State $Root \gets \{ \text{Status: "not-visited"}, \text{Patches: } Ps, \text{Edges: } \varnothing \}$
            \label{line:exec_scheduler:init}
        \While{there are "not-visited" nodes under $Root$}
                \label{line:exec_scheduler:mainloop}
            \State $Cur \gets Root$
            \State $S \gets \{\text{Loc: } Cb . \text{TestEntry}\}$
            \While{$S$ is not a termination state}
                \State $Pc \gets \{ P \in Cur . \text{Patches} \:|$ \newline
                    \hspace*{5em} $P [ S . \text{Loc} ] \neq Cb . \text{Stmt} [ S . \text{Loc} ] \}$
                \If{$Pc \neq \varnothing$}
                        \label{line:exec_scheduler:pc_neq_nothing}
                    \State $Cur, S \gets \textsc{EvalPatches}(Cur, S, Cur . \text{Patches})$
                        \label{line:exec_scheduler:call_eval_patches}
                \Else
                        \label{line:exec_scheduler:pc_eq_nothing}
                    \State $S \gets \textsc{Execute}(S, Cb . \text{Stmt} [ S . \text{Loc} ])$
                \EndIf
            \EndWhile
            \State $Cur . \text{Status} \gets \text{"test-finished"}$
                \label{line:exec_scheduler:update_test_finished}
            \State $\textsc{ReportResults}(Cur . \text{Patches},\: S)$
                \label{line:exec_scheduler:test_fin}
        \EndWhile
        
        \Statex
        
        \Procedure{EvalPatches}{$Cur, S, Ps$}
            \If{$Cur . \text{Status} = \text{"not-visited"}$}
                \State $Cur . \text{Status} \gets \text{"visited"}$
                    \label{line:exec_scheduler:first_visit}
                \For{$P \in Ps$}
                    \State $Ch \gets \textsc{Capture}(S, P [ S . \text{Loc} ])$
                    \If{$S . \text{Edges} [Ch]$ is defined}
                            \label{line:exec_scheduler:find_child}
                        \State $S . \text{Edges} [Ch] . \text{Patches} \gets \newline
                        \hspace*{5em} S . \text{Edges} [Ch] . \text{Patches} \cup \{ P \}$
                    \Else
                        \State $S . \text{Edges} [Ch] \gets \{ \newline
                            \hspace*{4.5em} \text{Status: "not-visited"}, \text{Patches: } \{ P \}, \text{Edges: } \varnothing \}$
                    \EndIf
                \EndFor
            \EndIf
            \State $Ch, Cur \gets \textsc{FindNotVisitedChild}(Cur . \text{Edges})$                 \label{line:exec_scheduler:FindNotVisitedSubtree}

            \State \Return $Cur, \textsc{Replay}(S, Ch)$
                \label{line:exec_scheduler:move_forward}
        \EndProcedure
    \end{algorithmic}
\end{algorithm}

The \napper is shown as \algoref{alg:exec_scheduler}. It takes as input the codebase of the project and a patch set to be validated. 
It maintains a variable $Root$, which corresponds to the root of the state-transition tree (\figref{fig:approach_state_transition_tree}). Each node in the tree is labeled with a status (``visited'', ``not-visited'', or ``test-finished''), a set of patches that belong to this node, and a mapping ``Edges'' from state changes to the child nodes. 
Initially at line~\ref{line:exec_scheduler:init}, the tree has only a root node labeled as ``not-visited'' with all patches belonging to it, indicating that nothing has been explored yet. Then, the loop body starting at line~\ref{line:exec_scheduler:mainloop} explores a path from the root node to a leaf node. The loop body resembles the plain patch validation, except that when there are some patches to the current statement ($Pc \neq \varnothing$, at line~\ref{line:exec_scheduler:pc_neq_nothing}), the \textsc{EvalPatches} procedure analyzes the state change of each patch, and chooses a child corresponding to a group of patches making the same state change to move forward. When a round of test execution finishes, test results for all patches belonging to the current node are reported (line~\ref{line:exec_scheduler:test_fin}).

\textsc{EvalPatches} is the critical procedure of the \napp. When the current node is visited for the first time (line~\ref{line:exec_scheduler:first_visit}), it captures the state change of each patch, and inserts a child node under the current node for each unique state change. For each patch, it searches for an existing edge corresponding to the state change $Ch$ of the patch (line \ref{line:exec_scheduler:find_child}): if there is such an edge, the patch merges into the patch set of the sub-node; otherwise, the patch forms a new sub-node on its own. Finally, \textsc{EvalPatches} finds a child whose subtree includes at least one ``not-visited'' node to continue execution (line~\ref{line:exec_scheduler:FindNotVisitedSubtree}), updating the current node and system state (line~\ref{line:exec_scheduler:move_forward}).

\subsection{Properties of the \NApper}

\begin{theorem_efficiency}
The number of rounds of test executions in \algoref{alg:exec_scheduler} is equal to the number of test-equivalence classes among patches, i.e., it removes all redundancies caused by test-equivalence.
\end{theorem_efficiency}

\begin{proof}[Proof Sketch]
Because two patches are test-equivalent if and only if they always make the same state change during the test execution, test-equivalent patches are never separated by \textsc{EvalPatches} into different sub-nodes, and non-test-equivalent patches must be separated by \textsc{EvalPatches} when they cause different state changes. Therefore, each leaf node labeled with ``test-finished'' corresponds to a test-equivalence class. We can see from \algoref{alg:exec_scheduler} that each round in the loop turns one label of a leaf node from ``not-visited'' to ``test-finished'' at line~\ref{line:exec_scheduler:update_test_finished}, so the number of rounds is equal to the number of ``test-finished'' leaf nodes, which is furtherly equal to the number of test-equivalence classes.
\end{proof}

\begin{theorem_soundness}
\algoref{alg:exec_scheduler} reports the same patch validation result as \algoref{alg:exec_plain}.
\end{theorem_soundness}

\begin{proof}[Proof Sketch]
First, we generalize \algoref{alg:exec_scheduler} so that the initial $S$ and $Root$ are read from inputs. Then we can prove that \algoref{alg:exec_scheduler} is equivalent to \algoref{alg:exec_plain} by induction on the depth of $Root$. \textbf{For the base case} when the depth is one (so $Root$ has no children, and thus \textsc{EvalPatches} is never executed), two algorithms are apparently equivalent because both algorithms execute the original program, leading to the same test result for all patches. \textbf{For the induction case}, consider an arbitrary patch $P$ that belongs to a leaf node $L$, which is reached by the path $Tree \to A \to ... \to L$. The test result for $P$ must be reported in the round that explores this path. During this round, let us consider the moment when $Cur$ changes from $Root$ to $A$, which is the moment when \textsc{EvalPatches} is called (line~\ref{line:exec_scheduler:call_eval_patches}) for the first time. Before this moment, both algorithms always execute the original program, so $S$ will be the same for the two algorithms up to this moment. Then, $S$ is updated to $S' \vcentcolon= \textsc{Execute}(S, Stmt)$ in \algoref{alg:exec_plain} and $S'' \vcentcolon= \textsc{Replay}(S, \textsc{Capture}(S, Stmt))$ in \algoref{alg:exec_scheduler}, where $Stmt \vcentcolon= P [ S . \text{Loc} ]$. Based on the accuracy requirement for \textsc{Capture} and \textsc{Replay}, we must have $S' = S''$. Then we can apply the induction hypothesis with the initial $S$ set to $S'$ and $Root$ set to $A$.
\end{proof}

\section{\EIns}
\label{section:approach_patch_virtualization}

The \napp algorithm can be implemented in different ways. To ensure the performance of the system, we implement this algorithm using instrumentation. That is, the tests and the patches are executed in the original runtime of the programming language, and only at certain locations, the instrumented code is invoked to detect equivalent patches and schedule executions. 

\begin{figure}[t]
    \centering
    \includegraphics[width=.9\columnwidth]{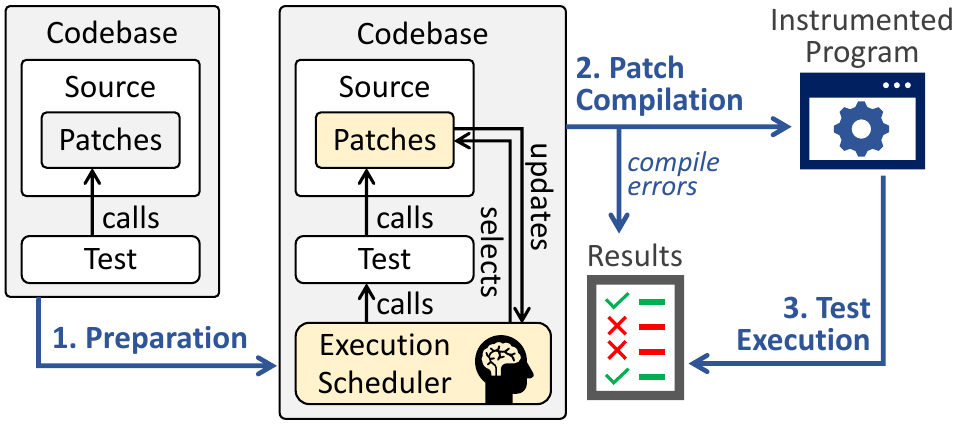}
    \caption{An overview of \ourtool}
    \label{fig:approach_overview}
\end{figure}

An overview of the patch validation process is shown in \figref{fig:approach_overview}. It takes three steps to validate all patches: preparation, patch compilation, and test execution.
In the first step, the codebase as well as the patches are instrumented to ensure the correct execution of the \napper. 
In the second step, we compile the instrumented codebase, dropping patches that are uncompilable. In the third step, the \napper iteratively runs the test suite, and stores the test result for each patch. 

Implementing the overall process of Algorithm~\ref{alg:exec_scheduler} with instrumentation is straightforward. First, a main procedure is injected into the codebase that initializes the state-transition tree and starts test executions. Then, at each patched location, the code is instrumented to evaluate the patches and update the tree. The main challenging part is how to implement the \textsc{Capture} and \textsc{Replay} in \algoref{alg:exec_scheduler}.

Existing mutation-testing approaches~\cite{just2011major,wong2018faster} instrument an interpreter at the mutated location for this purpose. However in APR, since a patch may invoke other methods and then execute a large chunk of code, using an interpreter is not only expensive but also difficult to implement. To solve this problem, we propose \eins. Our approach inserts code before and after the patched statement, such that the patch itself executes normally in the original runtime, and the inserted code is responsible for detecting what changes the patch has made to the system state and reverting the changes, i.e., the change is intercepted. 

In the preparation step, our approach adds instrumentation code around each patch, turning each patch into a \code{capture} component and a \code{replay} component. All such components are woven together into the codebase, effectively implementing mutant schemata. 
Finally, in the test execution step, the \napper calls the corresponding component as the \textsc{Capture} and \textsc{Replay} procedures.

In the rest of this section, we shall first introduce a small programming language, IMP+, for illustration. Then we describe how to intercept the changes based on IMP+. Finally, we show how to deal with uncompilable patches. 


\subsection{IMP+}

IMP+ is an enhancement of the classic IMP language~\cite{pierce2010software} with control flow constructs. IMP+ contains the commonalities between imperative programming languages, so we use it to illustrate our instrumentation process.
The syntax of IMP+ is shown below:

{\small
\newcolumntype{L}{>{$}l<{$}}
\newcolumntype{R}{>{$}r<{$}}
\setlength\tabcolsep{5pt}

\begin{supertabular}{LLLR}
    \\
    \tt s &\rightarrow& \tt s_1~s_2 & \textit{(statements)} \\
    \tt  &|& \tt x:=e;\\
    \tt  &|& \tt x:=m(e_1, \ldots, e_n);\\
    \tt  &|& \tt if(e)~s_1~else~s_2 \\
    \tt  &|& \tt try~s_1~catch(x)~s_2\\
    \tt  &|& \tt while(e)~s \\
    \tt  &|& \tt break; \\
    \tt  &|& \tt continue; \\
    \tt  &|& \tt return~e; \\
    \tt  &|& \tt throw~e;\\
    \tt  &|& \tt \ldots\\
    \\
    \tt e &\rightarrow& \tt e_1+e_2  & \textit{(expressions)} \\
    \tt  &|& \tt \ldots\\
    \\
\end{supertabular}
}

The runtime system state of an IMP+ program is a pair $\langle \sigma, \omega \rangle$, where the \textit{data state} $\sigma$ is a function mapping variables to their values, and the \textit{control state} $\omega$ can be one of the following values that reflects the effect of control flow statements:
\begin{itemize}
\item \code{Normal}, indicating that we should normally execute the next statement;
\item \code{Break}, indicating that a loop should break;
\item \code{Continue}, indicating that a loop should skip to the end of its body;
\item \code{Return $v$}, indicating that a method has returned $v$;
\item \code{Exception $e$}, indicating that an exception $e$ has been thrown.
\end{itemize}
The main operational semantic rules of IMP+ are shown in \figref{fig:approach_imp_semantic}, where \eval{\sigma}{e}{v} means that expression $e$ evaluates to $v$ under the system state $\langle \sigma, {\tt Normal} \rangle$, and \step{\sigma}{s}{\sigma'}{\omega'} means that statement $s$ changes the system state from $\langle \sigma, {\tt Normal} \rangle$ to $\langle \sigma', \omega' \rangle$. For simplicity, we only show the rules related to control flow change, and omit standard rules such as those for assignments and conditionals. 


Since a state includes two parts, we need to intercept the changes to both parts. Below, we discuss how to deal with changes to the control state and the data state respectively.

\subsection{Changes to the Control State}
\label{section:approach_control_state}
Recall that the \code{capture} component is responsible for detecting what changes a patched statement has made, and reverts the change; the \code{replay} component is responsible for replaying the recorded change. For changes to both states, the main challenge for implementing the two components lies in the control state, where the statement following the patched statement in the code file is not always the next statement to be executed at runtime. Therefore, we need a reliable way to detect and revert control state changes after the execution of the patch. Furthermore, since different programming languages have different language constructs for changing the control states, the design of these two components is by nature language-dependent.

To cope with the challenges, 
we propose a design process, \procref{proc:capture_control_state}, to systematically design \code{capture} and \code{replay} components for the control state $\omega$ given the semantic rules of the target programming languages. 
\xyaadd{While this process involves human effort, it only needs to be done once when migrating \ourtool to a new language: once these components are implemented, they are fully automatic and can be used for all patches under this language.} 

{
\floatname{algorithm}{Process}
\begin{algorithm}
    \newcounter{_alg_counter_bkp}
    \setcounter{_alg_counter_bkp}{\value{algorithm}}
    \setcounter{algorithm}{0}
    
    \caption{Capturing/replaying control state changes}
    \label{proc:capture_control_state}
    
    \setcounter{algorithm}{\value{_alg_counter_bkp}}
    
    \begin{algorithmic}[1]
        \Statex \textbf{Input:} Semantic rules $S$
        \Statex \textbf{Output:} Capture component $C$, replay component $R$
        \State $C \gets \lambda \tt s, {\tt s}$
        \State $R \gets {\tt skip}$
        \State $T \gets$ possible types of $\omega$ in $S$
        \While{$(T \setminus \{ {\tt Normal} \}) \neq \varnothing$}
            \State Choose a $t \in (T \setminus \{ {\tt Normal} \})$
            \State $T \gets T \setminus \{t\}$
            \State \#1: Pick an $R_c \in S$ that can change $\omega$ from $t$ to {\tt Normal}, and an $R_r \in S$ that can change $\omega$ from {\tt Normal} to $t$.
            \State \#2: Use $R_c$ to write a capture component $c_t(s)$ for any target statement $s$ and the type $t$.
            \State \#3: Refine $c_t(s)$ to ensure that it preserves the semantics of $s$ when it yields a state in $T$.
            \State \#4: Use $R_r$ to write a replay component $r_t$ for type $t$.

            \State $C \gets \lambda s, c_t(C(s))$
                \label{line:capture_control_state:stack_capture}
            \State $R \gets r_t ; R$
                \label{line:capture_control_state:stack_replay}
        \EndWhile
    \end{algorithmic}
\end{algorithm}
}

The generated \code{capture} component takes a patch as input, and produces a code snippet for executing the patch and capturing the change in variable {\tt change}. The generated \code{replay} is a code snippet replaying the change captured in {\tt change}. \procref{proc:capture_control_state} works by capturing and reverting each type of abnormal control state at a time. For each type, there are four manual steps (\#1 to \#4) to craft a \code{capture} component and a \code{replay} component for this exact type of control state. Then, these components are stacked together (line \ref{line:capture_control_state:stack_capture}-\ref{line:capture_control_state:stack_replay}) to cope with the whole control state. 

\begin{figure}[t]
    \centering

\newcommand{\rowspace}{\\[.75em]}
{\small
$$
\begin{array}{c}
\infer[\textsc{\scriptsize E-Break}]{\step{\sigma}{\tt break;}{\sigma}{\tt Break}
}{}
\rowspace
\infer[\textsc{\scriptsize E-Continue}]{\step{\sigma}{\tt continue;}{\sigma}{\tt Continue}
}{}
\rowspace
\infer[\textsc{\scriptsize E-Return}]{\step{\sigma}{\tt return~e;}{\sigma}{\tt Return~v}}{\eval{\sigma}{\tt e}{v}}
\rowspace
\infer[\textsc{\scriptsize E-Throw}]{\step{\sigma}{\tt throw~e;}{\sigma}{\tt Exception~v}}{\eval{\sigma}{\tt e}{v}}
\rowspace
\infer[\textsc{\scriptsize E-Seq}]{\step{\sigma}{\tt s_1~s_2}{\sigma''}{\omega''}}
{\step{\sigma}{\tt s_1}{\sigma'}{{\tt Normal}} & 
\step{\sigma'}{\tt s_2}{\sigma''}{\omega''}}
\rowspace
\infer[\textsc{\scriptsize E-SeqSkip}]{\step{\sigma}{\tt s_1~s_2}{\sigma'}{\omega'}}{\step{\sigma}{\tt s_1}{\sigma'}{\omega'} & \omega' \neq {\tt Normal}}
\rowspace
\infer[\textsc{\scriptsize E-While}]{\step{\sigma}{\tt while(e)~s}{\sigma''}{\omega''}}{
\begin{array}{cc}
 \eval{\sigma}{\tt e}{\tt true} &
 \step{\sigma}{\tt s}{\sigma'}{\omega'} \\
 \omega' \in \{ {\tt Normal}, {\tt Continue} \} &
 \step{\sigma'}{\tt while(e)~s}{\sigma''}{\omega''}
\end{array}
}
\rowspace
\infer[\textsc{\scriptsize E-WhileFalse}]{\step{\sigma}{\tt while(e)~s}{\sigma}{\tt Normal}}{
 \eval{\sigma}{\tt e}{\tt false} 
}
\rowspace
\infer[\textsc{\scriptsize E-WhileBreak}]{\step{\sigma}{\tt while(e)~s}{\sigma'}{\tt Normal}}{
 \eval{\sigma}{\tt e}{\tt true} &
 \step{\sigma}{\tt s}{\sigma'}{\tt Break} 
}
\rowspace
\infer[\textsc{\scriptsize E-WhileSkip}]{\step{\sigma}{\tt while(e)~s}{\sigma'}{\omega'}}{
\begin{array}{cc}
 \eval{\sigma}{\tt e}{\tt true} &
 \step{\sigma}{\tt s}{\sigma'}{\omega'} \\
 \multicolumn{2}{c}{\omega' \in \{ {\tt Return}~v, {\tt Exception}~e \}}
\end{array}
}
\rowspace
\infer[\textsc{\scriptsize E-Call}]{\step{\sigma}{\tt x:=m(e_1, \ldots, e_n)}{\sigma'[x\backslash v]}{\tt Normal}}{
    \begin{array}{l}
  {\it body}({\tt m})={\tt s} \qquad \eval{\sigma}{\tt e_i}{v_i} \\
  \step{\sigma[{\it para}({\tt m},i)\backslash v_i]}{\tt s}{\sigma'}{{\tt Return}~v} 
    \end{array}
}
\rowspace
\infer[\textsc{\scriptsize E-CallExc}]{\step{\sigma}{\tt x:=m(e_1, \ldots, e_n)}{\sigma'}{{\tt Exception}~e}}{
    \begin{array}{l}
  {\it body}({\tt m})={\tt s} \qquad \eval{\sigma}{\tt e_i}{v_i} \\
  \step{\sigma[{\it para}({\tt m},i)\backslash v_i]}{\tt s}{\sigma'}{{\tt Exception}~e}
    \end{array}
}
\rowspace
\infer[\textsc{\scriptsize E-Catch}]{\step{\sigma}{\tt try~s_1~catch(x)~s_2}{\sigma''}{\omega''}}{
  \step{\sigma}{\tt s_1}{\sigma'}{{\tt Exception}~e} & \step{\sigma'[x\backslash v]}{\tt s_2}{\sigma''}{\omega''}
}
\rowspace
\infer[\textsc{\scriptsize E-CatchSkip}]{\step{\sigma}{\tt try~s_1~catch(x)~s_2}{\sigma'}{\omega'}}{
  \step{\sigma}{\tt s_1}{\sigma'}{\omega'} & (\omega' \neq {\tt Exception}~e)
}
\end{array}
$$
}
    \caption{Semantic rules of IMP+}
    \label{fig:approach_imp_semantic}
\end{figure}

We illustrate this process with the semantic rules of IMP+ as shown in \figref{fig:approach_imp_semantic}, where \code{Break}, \code{Continue}, \code{Return $v$} and \code{Throw $e$} are abnormal control states that we concern.

\textbf{For the \code{Break} state:} At step \#1, we find that the only possible $R_c$ is \textsc{E-WhileBreak} and the only possible $R_r$ is \textsc{E-Break}. At step \#2, we use this rule to write a \code{capture} component that turns a \code{Break} state into \code{Normal}:

\vspace{.5em}
\begin{addmargin}[2em]{0em}
\small\tt

while(true) \{s\} \\
change := "break";
\end{addmargin}
\vspace{.5em}

Among possible states in $T = \{\allowbreak \code{Normal},\allowbreak \code{Continue},\allowbreak \code{Return}~ v,\allowbreak \code{Exception}~e \}$, the component above does not preserve the semantics of the former two states: if the patch ends up with a \code{Normal} or \code{Continue} state, it causes an endless loop. Therefore at step \#3, We modify the \code{capture} component to fix these problems:

\vspace{.5em}
\begin{addmargin}[2em]{0em}
\small\tt

flag := 0; \\
while(true) \{
\begin{addmargin}[1em]{0em}
flag := flag+1; \\
if(flag>1) break; \\
\{s\} \\
flag := flag+1;\}
\end{addmargin}
if(flag=1) change := "break";\\
if(flag=2) continue;
\end{addmargin}
\vspace{.5em}

The added \code{flag} variable will distinguish \code{Break} from \code{Normal} and \code{Continue}, so that it correctly preserves the semantics of \code{s} when it does not cause a \code{Break} state. Finally at step \#4, we use \textsc{E-Break} to write a \code{replay} component:

\vspace{.5em}
\begin{addmargin}[2em]{0em}
\small\tt

if(change="break") break;
\end{addmargin}
\vspace{.5em}

\textbf{For the \code{Continue} state:} At step \#1, we find $R_c=$ \textsc{E-While} and $R_r=$ \textsc{E-Continue}. At step \#2\textendash\#3, the \code{capture} component is mostly the same as the component for \code{Break}, but the last line should be changed to \code{if(flag=2) change:="continue";}. At step \#4, the \code{replay} component is \code{if(change="continue") continue;}.

\textbf{For the \code{Return $v$} state:} At step \#1, $R_c=$ \textsc{E-Call} and $R_r=$ \textsc{E-Return}. At step \#2, the \code{capture} component wraps the statement into a new method and then calls it:

\vspace{.5em}
\begin{addmargin}[2em]{0em}
\small\tt

def m() \{s\} \\
x := m(); \\
change := "return"; retval := x;
\end{addmargin}
\vspace{.5em}

For each state in $T = \{\allowbreak \code{Normal},\allowbreak \code{Exception}~e \}$, this component fails when the patch ends up with a \code{Normal} state. At step \#3, we fix it by \code{return}ing a unique special value at the end of the wrapping method:

\vspace{.5em}
\begin{addmargin}[2em]{0em}
\small\tt

def m() \{\{s\} return "--special--";\} \\
x := m(); \\
if(x!="--special--") \{
\begin{addmargin}[1em]{0em}
change := "return"; retval := x;\}
\end{addmargin}
\end{addmargin}
\vspace{.5em}

At step \#4, we come up with this \code{replay} component:

\vspace{.5em}
\begin{addmargin}[2em]{0em}
\small\tt

if(change="return") return retval;
\end{addmargin}
\vspace{.5em}

\textbf{For the \code{Exception $e$} state,} we similarly use $R_c=$ \textsc{E-Catch} to write the \code{capture} component as \code{try\allowbreak\{s\} catch(e) \{change:="exc"; exc:=e;\}}, and uses $R_r=$ \textsc{E-Throw} to write the \code{replay} component as \code{if\allowbreak(change="exc") throw exc;}.

In the end, we get the following components that capture and replay all control state changes. 
\begin{align*}
C(s) =&\: c_\text{Exception} ( c_\text{Return} ( c_\text{Continue} ( c_\text{Break} (s) ) ) ) \\
R =&\: r_\text{Exception}; r_\text{Return}; r_\text{Continue}; r_\text{Break}; \tt skip
\end{align*}



\subsection{Changes to the Data State}

After the control state is intercepted in \procref{proc:capture_control_state}, further intercepting changes to the data state is easier because we can detect and revert the modified variables in the next statement following the patch -- the next statement is now guaranteed to run. On the other hand, unlike the control state where reverting always sets it to \code{Normal}, here we need an efficient way to detect and revert the data state to its version before executing the patched statement. Simply recording the whole data state is not feasible due to its size. 

To cope with this challenge, we use lightweight static analysis to detect the change scope that each patch may bring to the data state, and record only the values within the scope. After executing the patched statement, the new values in the scope can be compared with the recorded ones for detecting and reverting the change. We further bypass the equivalence detection on patches whose change scope is too large: these patches are validated sequentially using the plain method but not with the \napper.


Our approach does not restrict the choice of the static analysis and the definition of the change scope, and the implementation could use any static analysis algorithm that best fits the target programming language and the APR tools. In the following, we describe an implementation of the two components for the IMP+ language, which is close to those used in \ourtool.

In IMP+, the only two types of statement that can change the data state are \code{$\tt x:=e$} and \code{$\tt x:=m(e_1, \ldots, e_n)$}, which change the value of a variable \code{x} to an expression or the return value of a method. As a result, we can define the change scope as a set of variables, and use an inter-procedural static analysis to produce this change scope. This static analysis can be easily implemented by scanning the code and collecting the left variables in all assignments. When a method call is encountered, all variables changed by the target method are also iteratively added until a fixed point is reached. 

Given a set of variables \code{$\{\tt x_1, \ldots x_n\}$} returned by the static analysis, we can refine the \code{capture} component in the following way to intercept data state changes.


\vspace{.5em}
\begin{addmargin}[2em]{0em}
\small

\code{$\tt old_1 := x_1;\: \ldots;\: old_n := x_n;$}\\
\code{\textit{(original capture component by \procref{proc:capture_control_state})}}\\
\code{$\tt if (old_1 \neq x_1)\: \{change[0] := x_1; \: x_1 := old_1;\}$}\\
\code{$\tt else\: \{change[0] := null;\}$}\\
\ldots \\
\code{$\tt if (old_n \neq x_n)\: \{change[n-1] := x_n;\: x_n := old_n;\}$}\\
\code{$\tt else\: \{change[n-1] := null;\}$}
\end{addmargin}
\vspace{.5em}

We similarly refine the \code{replay} component.

\vspace{.5em}
\begin{addmargin}[2em]{0em}
\small

\code{$\tt if (change[0]\neq null)\: x_1 := change[0];$} \\
\ldots\\
\code{$\tt if (change[n-1]\neq null)\: x_n := change[n-1];$} \\
\code{\textit{(original replay component by \procref{proc:capture_control_state})}}
\end{addmargin}
\vspace{.5em}

This implementation still has two issues in efficiency. First, the change scope may include too many variables, causing excessive overhead at runtime. To deal with this issue, we simply use a purity analysis~\cite{bravenboer2009strictly} rather than an inter-procedural change scope analysis. A method is \emph{pure} if it does not change the data state. A purity analysis involves only a Boolean result and can be efficiently conducted. If a patched statement calls any impure method, we consider its change scope too large and it will bypass the equivalence detection.

Second, a variable may store a large data structure where copying and comparison may be costly. To deal with this issue, we keep an allowlist of types that are either primitive types or compound types of a small, fixed size. If the change scope includes any variable whose type is not in the allowlist, we consider the change scope too large and the patch will bypass the equivalence detection.


\subsection{Properties of Interception-based Instrumentation}

\begin{theorem_efficiency}
The time cost of the designed \code{capture} and \code{replay} components in the above example is at most linear to the syntactic size of the patched statement,
i.e., the overhead of \eins is small even if the patched statement may execute for a long time (e.g., contains a loop or a recursive call).
\end{theorem_efficiency}

\begin{proof}[Proof Sketch]
The final \code{capture} and \code{replay} components consist of statements that deal with changes to the control state and the data state. For the control state, a constant number of statements are added to the component. For the data state, the number of added statements is linear to the number of possibly modified variables, which is no more than the syntactic size of the statement.
\end{proof}

\begin{theorem_soundness}
$\textsc{Replay}(S, \textsc{Capture}(S, T)) = \textsc{Execute}(S, T)$ for any state $S$ and any statement $T$, i.e., calling the \code{capture} component and then the \code{replay} component for any patch is equivalent to executing the patch.
\end{theorem_soundness}

\begin{proof}[Proof Sketch]
The construction of both components is incremental, i.e., we build a component for each aspect of the state change, and combine them together. Each individual component is sound within its own scope by design, and the combining process is also sound: when combining components for control state changes, step \#3 in \procref{proc:capture_control_state} ensures that an added component does not break previous components; when refining the component for data state changes, the control state is already reverted, so that the refined component always executes normally regardless of control flow effects in the patch.
\end{proof}

\subsection{Compilation Isolation}

Unlike mutation testing that assumes mutants always compile, patches generated by an APR approach may not compile, leading to a challenge when all patches are woven into one program: if any patch cannot compile, the whole program cannot compile. It is impossible to detect and remove uncompilable patches by individually compiling each patch, because it will nullify the acceleration effect of mutant schemata.

An intuition to solve this challenge is to rely on the error messages from the compiler: if the error messages could precisely pinpoint all uncompilable code snippets, 
we can compile the codebase once and then remove the uncompilable patches based on the error messages.

However, the error messages are often not perfectly 
precise: for example, a patch in the middle of a method may lead to compile errors at the end of this method, and it is often the case that an error stops the compilation of the rest of the method. Therefore, we cannot rely on the error messages to precisely pinpoint all uncompilable patches. 

To solve this problem, we propose the concept \textit{isolation unit} to measure the granularity of the preciseness in the error messages of a compiler: a compilation error within an isolation unit will only cause error messages within the isolation unit but not affect the compilation of other parts of the code. An isolation unit must exist for any compiler: in the extreme case, the whole codebase is the isolation unit. However, our observation is that modern compilers usually have more fine-grained isolation units. For example, many compilers compile each file individually into an object file (\code{.java} files into Java bytecode files, and \code{.c} files into assembly files), and thus each file is an isolation unit. Furthermore, a method or a procedure in many imperative programming languages, including Java and C, is also an isolation unit, because these programming languages are carefully designed such that compilers only need to perform intra-procedural analysis when compiling the code (inter-procedural analysis is often applied at a much later stage for optimizing the code where the compilation errors have already been detected).


Therefore, to isolate compile errors, our interception-based instrumentation approach should wrap each patch in an isolation unit. 
Let us assume that methods are also isolation units for an IMP+ compiler. Each \code{capture} component produced in \secref{section:approach_control_state} is already in an individual isolation unit because we wrap the patch in a method to deal with the \code{Return $v$} control state.
Then we can compile the codebase woven with all patches with two rounds of compilation: the first round identifies all uncompilable patches from error messages, which will be removed, and the second round compiles the codebase with only the remaining patches.

\section{Limitation}
\label{section:limitation}

While our approach is theoretically sound and can be applied to arbitrary patches \xyaadd{under our idealized problem definition in \secref{section:approach_test_def}, there are two practical limitations when applying it in the real world}. We will evaluate them in our experiment.

\textbf{Patch Limitation:} We define a patch as a modification of statements. Therefore, patches that modify other parts in the program (e.g., the definition of a field) are out of the scope of \ourtool. We detect this case in the preparation step, and fall back to plain validation for affected patches.
    
\textbf{Test Limitation:} We define executing the test as stepping through a state machine, where the state-transition should be \emph{stable}, i.e., being exactly the same across multiple runs. In practice, most test cases are stable -- otherwise, it would be difficult to troubleshoot a test failure. 
We perform runtime checks in \algoref{alg:exec_scheduler} to detect rare unstable test cases\footnote{In the \textsc{EvalPatches} procedure and when the test ends, we assert that the observed system state should be consistent with the record on the state-transition tree. For example, if the test execution finishes but $Cur . \text{Status} = \text{"visited"}$, the test case is considered unstable.}, and fall back to plain validation for them.

\xyaadd{Also note that \ourtool is only suitable for test-based generate-and-validate APR. While most existing studies focus on this category, many other kinds of defects, such as alerts by static analyzers~\cite{bessey2010few} or anomalies in logs~\cite{zte1,zte2}, may also be targets of APR. 
\ourtool is not designed for these approaches because they do not run tests for patch validation.}

\section{Implementation}
\label{section:approach_impl}

After the challenges for adapting mutant schemata and mutant deduplication are solved, we build \ourtool, a general-purpose patch validator for Java, which adapts all five classes of acceleration techniques suitable for patch validation. We choose Java as the target programming language because it is supported by most APR tools\footnote{\href{https://program-repair.org/}{program-repair.org} lists 23 APR tools for repairing Java programs, which is the largest among all programming languages.}.

To achieve mutant schemata and mutant deduplication, we implement the \napp algorithm described in \secref{section:approach} and the interception-based instrumentation by following the design process described in \secref{section:approach_patch_virtualization}.

The \code{capture/replay} component for Java is similar to the IMP+ example as described in \secref{section:approach_patch_virtualization}, because both languages share similar features. For the control state, control flow statements in Java can only be \code{break}, \code{continue}, \code{return}, or \code{throw}~\cite{jls8}, which are all covered by IMP+. For the data state, Java programs similarly change the data state by assignment statements. 
The \code{capture/replay} component for Java has three differences compared against IMP+: 1) Java allows to assign a label to a loop and jumps out of a labeled loop. To support this syntax, we detect labeled \code{break/continue} statements in the preparation step, and add the same label in the generated \code{replay} component. 2) Java methods can declare a list of checked exceptions to be thrown. Generated components should inherit this list from the original method. 3) A patch may access local variables in the original method and fields in the original class. To make it possible, we put generated methods in the same class as the original patch, and declare auxiliary fields for passing around local variables between the original method and the generated methods.

Adapting the three other acceleration techniques is straightforward:
\textbf{Test virtualization} is adapted by resetting global states (except for the state-transition tree that should be shared across runs) before each round of test execution. Current \ourtool implementation uses an existing library, VMVM~\cite{bell2014unit}, for test virtualization.
\textbf{Test case prioritization} is adapted by sorting test cases in the test suite. Following the two heuristics mentioned in \secref{section:survey_test_prio}, \ourtool first executes the failing test cases, then other test cases in the same package of a patched location, and then test cases in other packages.
\textbf{Parallelization} is adapted by spawning multiple instances of the patch validator with a process pool, one instance dealing with a subset of patches, each subset corresponding to all patches to the same fault location as generated by the APR approach.

\section{Experiment Setup}
\label{section:exp_setup}

\subsection{Research Questions}

Our evaluation aims to answer these research questions.

\begin{enumerate}
    \item[RQ1.] \textbf{Overall performance.}
    How fast patch validation could be with \ourtool?
    \item[RQ2.] \textbf{Technique effectiveness.} Does each technique in \ourtool contribute to the final performance?
    \item[RQ3.] \textbf{Feasibility.} Will \ourtool fail or report incorrect validation results, affecting the ability of APR?
    \item[RQ4.] \xyaadd{\textbf{Generalizability.} How does the speed and feasibility of \ourtool generalize to different kinds of programs?}
\end{enumerate}

\subsection{The Benchmark}


We used Defects4J~\cite{just2014defects4j} v1.2 as the benchmark for RQ1 through RQ3.
Defects4J v1.2 is a widely used APR benchmark on Java, containing 395 bugs in open-source Java projects.
We chose four publicly available APR tools covering different kinds of APR approaches to be studied in this evaluation:

\begin{itemize}
    \item Recoder~\cite{zhu2021syntaxguided}, a deep-learning-based tool fixing 53 bugs in the benchmark.
    \item TBar~\cite{liu2019tbar}, a template-based tool fixing 41 bugs.
    \item SimFix~\cite{jiang2018shaping}, a heuristic-based tool fixing 33 bugs.
    \item Hanabi~\cite{xiong2021l2s}, a decision-tree-based tool fixing 27 bugs.
\end{itemize}

We obtained the replication package for each tool, and collected all generated candidate patches under the original time-out value for evaluation.

We did not conduct the evaluation on all 395 bugs in Defects4J because the computational cost would be unacceptable to execute multiple baselines for the patches generated by the four APR tools. 
Instead, we sampled the dataset in two possible ways: 1) for each studied APR tool, choosing all bugs it can fix as reported by its authors (the \textit{fixable} dataset); 2) choosing 30 bugs randomly from the whole dataset (the \textit{random} dataset). 
The fixable dataset helps us understand whether APR can be accelerated: if the time used in each fixable bug is shorter, we can use a shorter time limit without affecting the effectiveness of the tool.
The random dataset is supplemented to avoid the threat of possible selection bias on the fixable dataset. Results on both datasets are reported.

\begin{table}[t]
    \centering\footnotesize
    \caption{Statistics of evaluated patches}
    \label{tab:expsetup_patch_count}
    \begin{tabular}{l@{\hspace{1.25em}}l@{\hspace{1.25em}}r@{\hspace{1.25em}}r@{\hspace{1.25em}}r@{\hspace{1.25em}}r}
        \toprule
        RQ & APR      & \#Bugs & \#PatchSets & \#Patches & Avg $\frac{\text{\#Patches}}{\text{Patch Set}}$ \\
        \midrule
         & Recoder  & 53      & 3096         & 101543     & 32.8 \\
        RQ1$\sim$3 & TBar     & 41      & 3138         & 110882    & 35.3 \\
        (fixable) & SimFix   & 33      & 1799         & 120788    & 67.1 \\
         & Hanabi   & 27      & 959          & 99627    & 103.9 \\
        \midrule
         & Recoder  & * 29      & 1956         & 34722     & 17.8 \\
        RQ1$\sim$3 & TBar     & * 29      & 4602         & 148140    & 32.2 \\
        (random) & SimFix   & 30      & 3161         & 227041    & 71.8 \\
         & Hanabi   & * 29      & 650          & 87990    & 135.4 \\
        \midrule
        RQ4 & Recoder  & 30      & 866         & 36565     & 42.2 \\
        \bottomrule
    \end{tabular}

    \vspace{.5em}
    {\raggedright\footnotesize
        *: Among the 30 selected random bugs, Recoder and Hanabi fails to run on Lang-25, and TBar fails to run on Mockito-20.
    }
\end{table}

\xyaadd{
In RQ4, we additionally experimented with Recoder under 30 random bugs in the IntroClassJava~\cite{durieux:hal-01272126} benchmark, as a supplement to Defects4J. This benchmark includes 297 small and buggy student assignments, which allows us to understand the generalizability of \ourtool.
}

\xyaadd{The statistics of evaluated patches is shown in \tabref{tab:expsetup_patch_count}}. The current evaluation costs over 18 months of single-CPU-core time in total, which is of the largest scale among all existing studies on accelerating APR within our knowledge.

\subsection{Experiment Methodology}

\textbf{In RQ1,} the patch validation time per bug using \ourtool is compared with the following two baselines:

\begin{itemize}
    \item \textbf{Plain.}
        Each patch is compiled using the \code{defects4j compile} command that ships with the Defects4J dataset, which uses an Ant script to compile changed files and perform necessary user-defined actions. Compilable patches are tested using the \code{defects4j test} command. This reflects the normal practice of evaluating APR techniques.
    \item \textbf{State-of-the-Art.}
        UniAPR~\cite{chen2021fast}, the current state-of-the-art patch validator on Java, is used to validate all patches. It incorporates test virtualization, test case prioritization, and parallelization, but not mutant schemata and mutant deduplication, the two techniques brought to patch validation by \ourtool for the first time. When compiling the patches, we use the OpenJDK compiler (the \code{javac} command) to compile only the patched file.  
\end{itemize}

To simulate the degree of parallelization on average hardware, the \xyaadd{patch validation job for each defect} is distributed to eight CPU cores for every baseline. \xyaadd{In other words, parallelization is already included in all the baselines. To understand the performance without parallelization, we also report the time needed when we perform serial validation in the {Plain} baseline. However, for a fair comparison, we do not treat the serial validation as a separate baseline, because otherwise the acceleration ratio can be easily manipulated: the more CPU cores we use in our experiment, the higher the acceleration ratio for the parallel validation.}

\xyaadd{We also measure the memory footprint of each approach during the experiment.}

\textbf{In RQ2,} we discuss the individual effectiveness of each acceleration technique in \ourtool. For Parallelization, its effectiveness is straightforward (roughly $N$ times acceleration when running on $N$ cores). For the other four techniques, namely Mutant Schemata (MS), Mutant Deduplication (MD), Test Virtualization (TV), and Test Case Prioritization (TCP), we empirically measure their effectiveness.

Among these four techniques, Mutant Schemata accelerates the patch compilation step, and the other three techniques accelerate the test execution step. Besides the original ExpressAPR implementation, we add intermediate configurations by removing each technique from \ourtool one by one, and measure their time usage. For the patch compilation step, we add one intermediate configuration (``-MS''); for the test execution step, we add three intermediate configurations (``-TCP'', ``-TCP -MD'', ``-TCP -MD -TV''). In this way, we can understand the effectiveness of the removed technique between two adjacent configurations. Note that this is not an ablation study due to implementation-level dependencies -- it can be hard to individually remove a technique from \ourtool\footnote{For example, Mutant Deduplication requires the state-transition tree to persist across executions, which is implemented as part of the Test Virtualization technique. Therefore, we cannot remove TV without breaking MD.}.


\textbf{In RQ3,} we compare the validation result (``plausible'', ``implausible'' or ``fails to validate'') of \ourtool against the Plain baseline. The \ourtool implementation may fail to accelerate the validation of a patch if it detects a violation of two limitations described in \secref{section:limitation}. It may also produce an incorrect result if it fails to detect such a violation that affects the validation result.
In this RQ, we evaluate how often these cases happen. We count the number of patches that \ourtool fails to validate or reports incorrect results (reported ``plausible'' when should be ``implausible'' or vice-versa), and analyze the reason.

\xyaadd{
\textbf{In RQ4,} we compare the validation time and the result of \ourtool against the patch validation command under the IntroClassJava dataset (similar to the Plain baseline). This helps us to understand how much the result of previous RQs generalize to small programs that may not have many redundancies. Due to the computational cost, We do not experiment with other repair tools or other baselines.
}

\subsection{Experiment Settings}
\label{section:exp_setup_implementation}

The experiment is run on a Xeon\textsuperscript{\textregistered} 8270 CPU server with eight processes, each using one dedicated CPU core.
For patches \ourtool ``fails to validate'', we fall back to the State-of-the-Art baseline in RQ1.
To deal with candidate patches with dead loops, following an existing study~\cite{ghanbari2020prf}, we set the test timeout to 5 seconds plus 1.5 times over the original test execution time.
UniAPR and \ourtool have offline procedures to measure the original test execution time or to analyze the purity of methods \xyaadd{(the analysis averagely takes 130 seconds)}, and they are excluded from time usage.

\section{Result Analysis}
\label{section:exp_result}

\newtcolorbox{rqbox}{left=1pt,right=1pt,top=1pt,bottom=1pt}

    \subsection{RQ1: Overall Performance}

The total patch validation time per bug in both datasets using each approach is shown in \figref{fig:result_overall_scatter}. \xyaadd{The Y-Axis is logarithmic due to the huge difference across baselines.} Note that UniAPR fails to run on 24\% of bugs, hence some points in the SOTA series are missing in the figure.
This result brings the following findings for the APR community:

\begin{figure}[t]
    \centering

    \begin{subfigure}[b]{\columnwidth}
        \includegraphics[width=\columnwidth]{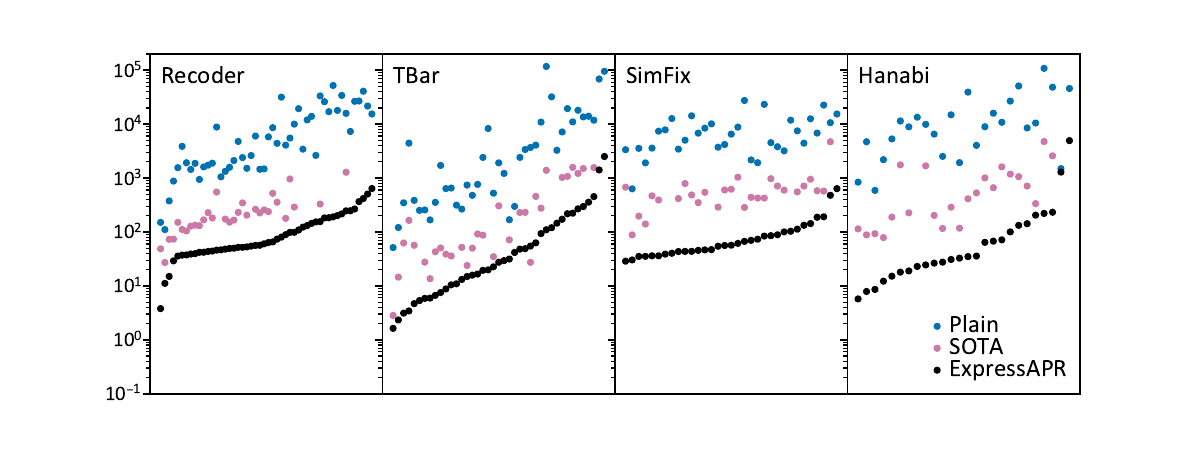}
        \vspace{-1.75em}
        \caption{fixable bugs}
    \end{subfigure}
    \begin{subfigure}[b]{\columnwidth}
        \includegraphics[width=\columnwidth]{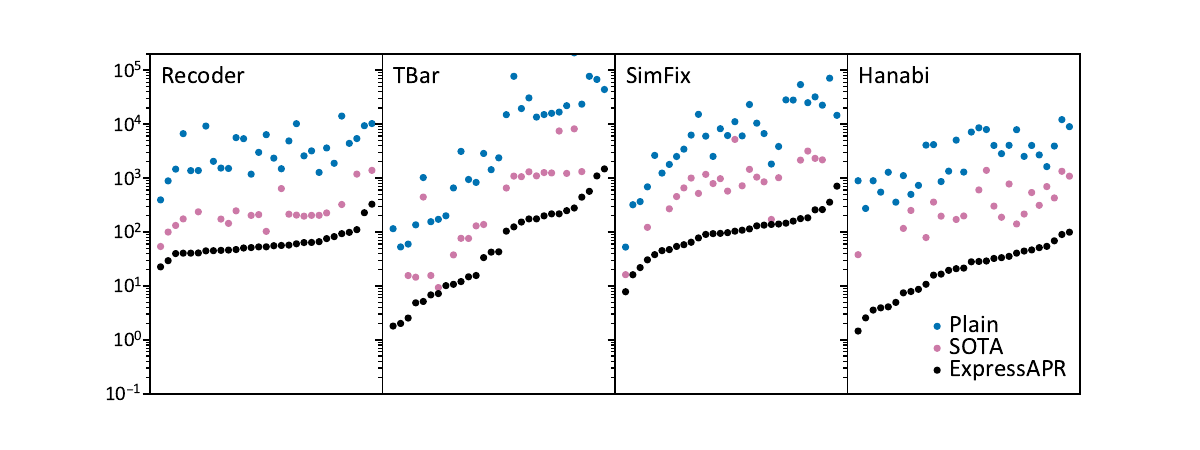}
        \vspace{-1.75em}
        \caption{random bugs}
    \end{subfigure}

    \caption{Patch validation time (seconds) per bug, \xyaadd{on a logarithmic scale}}
    \label{fig:result_overall_scatter}
\end{figure}

\textbf{1. \ourtool accelerates patch validation to a new degree in a variety of settings.} From \figref{fig:result_overall_scatter} we can see that in every studied APR tool, patch validation with \ourtool consumes significantly less time compared with the two baselines for almost all bugs. In the fixable dataset, \ourtool shows an acceleration of \accOverPlain{}x over the Plain baseline, or \accOverSota{}x over the State-of-the-Art baseline on bugs where UniAPR successfully runs.
In the random dataset, \ourtool shows a similar acceleration of 108.9x over Plain or 10.3x over State-of-the-Art. This indicates that the performance of ExpressAPR generalizes to different APR tools and different bugs.

\begin{figure}[t]
    \centering

    \begin{subfigure}[b]{\columnwidth}
        \includegraphics[width=\columnwidth]{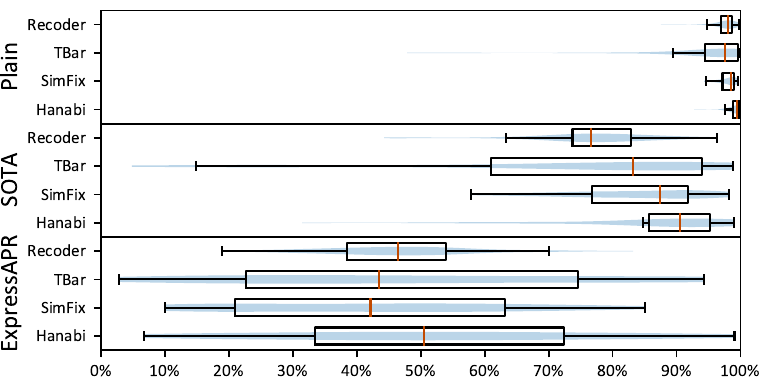}
        \vspace{-1.5em}
        \caption{fixable bugs}
    \end{subfigure}
    \begin{subfigure}[b]{\columnwidth}
        \includegraphics[width=\columnwidth]{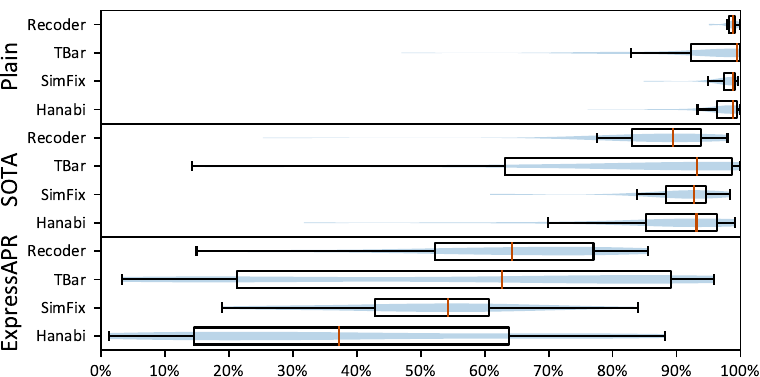}
        \vspace{-1.5em}
        \caption{random bugs}
    \end{subfigure}

    \caption{Percentage of time spent for patch validation}
    \label{fig:result_overall_pvratio}
\end{figure}

\textbf{2. Patch validation is no longer the speed bottleneck of APR if mutation testing techniques are systematically adapted.} The percentage of time spent for patch validation when repairing each bug in the fixable dataset is shown in \figref{fig:result_overall_pvratio}. When using the stock Defects4J command (the Plain baseline), patch validation takes more than 95\% of the repair time. If the previous best patch validation approach (the State-of-the-Art baseline) is used, this portion is reduced to $75\% \sim 95\%$, still the bottleneck. \ourtool has reduced this portion to about 50\%, so patch validation is now as fast as patch generation. This suggests that future APR acceleration work should also consider the patch generation phase.

\begin{figure}[t]
    \centering
    \includegraphics[width=.95\columnwidth]{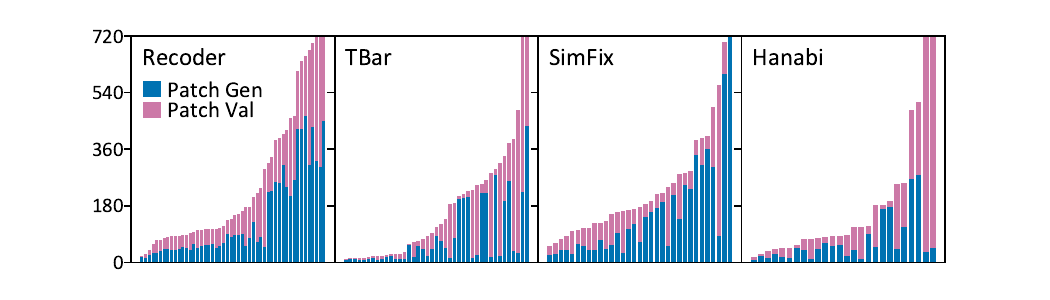}
    \caption{Total repair time (seconds) of fixable bugs using \ourtool}
    \label{fig:result_overall_eachtime}
\end{figure}

\textbf{3. The speed of APR satisfies the expectation of most users.} 
In our experiment with an 8-core configuration, when \ourtool is used, the total repair time for the fixable dataset has greatly reduced to less than 3 minutes for more than half bugs and less than 10 minutes for nearly all bugs, as shown in \figref{fig:result_overall_eachtime}. 
This execution time satisfies the expectation of most users in an existing survey~\cite{noller2022trust}. 

\begin{figure}[t]
    \centering
    \includegraphics[width=.95\columnwidth]{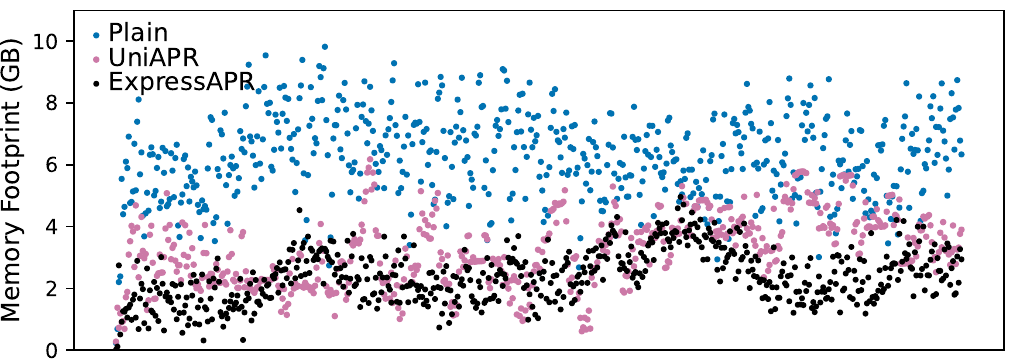}
    \caption{Memory footprint of patch validation over one hour}
    \label{fig:result_mem_footprint}
\end{figure}

\xyaadd{
\textbf{4. The memory overhead of \ourtool is negligible.} While \ourtool theoretically has a memory overhead for the State-Transition Tree, we do not observe additional memory usage in the experiment. In fact, \ourtool uses much less memory compared with Plain, because it avoids heavy frameworks (Maven or Ant) when running tests. \figref{fig:result_mem_footprint} shows the memory footprint of all three approaches over the first hour of the experiment, which validates the candidate patches in the fixable dataset in random order.
}

\xyaadd{
\textbf{5. The acceleration ratio from parallelization is close to the number of CPU cores used.}
To understand the performance without parallelization, we further randomly sampled 300 patch sets from the fixable dataset, and validated them with only one process on a single CPU core (in contrast to eight processes each using one core). The time usage becomes 7.2 times as long as Plain, which is close to the number of CPU cores (8). This is intuitive because the validation of patches at different locations has no dependency on each other and is thus naturally parallelizable.
}

\subsection{RQ2: Technique Effectiveness}

In this RQ, we separately measure the effectiveness of four acceleration techniques in its step (patch compilation for MS; test execution for MD, TC, and TCP). Results on both datasets are shown in \figref{fig:result_speed_plot}. Each point $(x\%, y)$ in the plot indicates that the time usage of this step among $x$\% patch sets is within $y$ seconds.
We have the following findings:

\begin{figure}[t]
    \centering

    \begin{subfigure}[b]{\columnwidth}
        \includegraphics[width=\columnwidth]{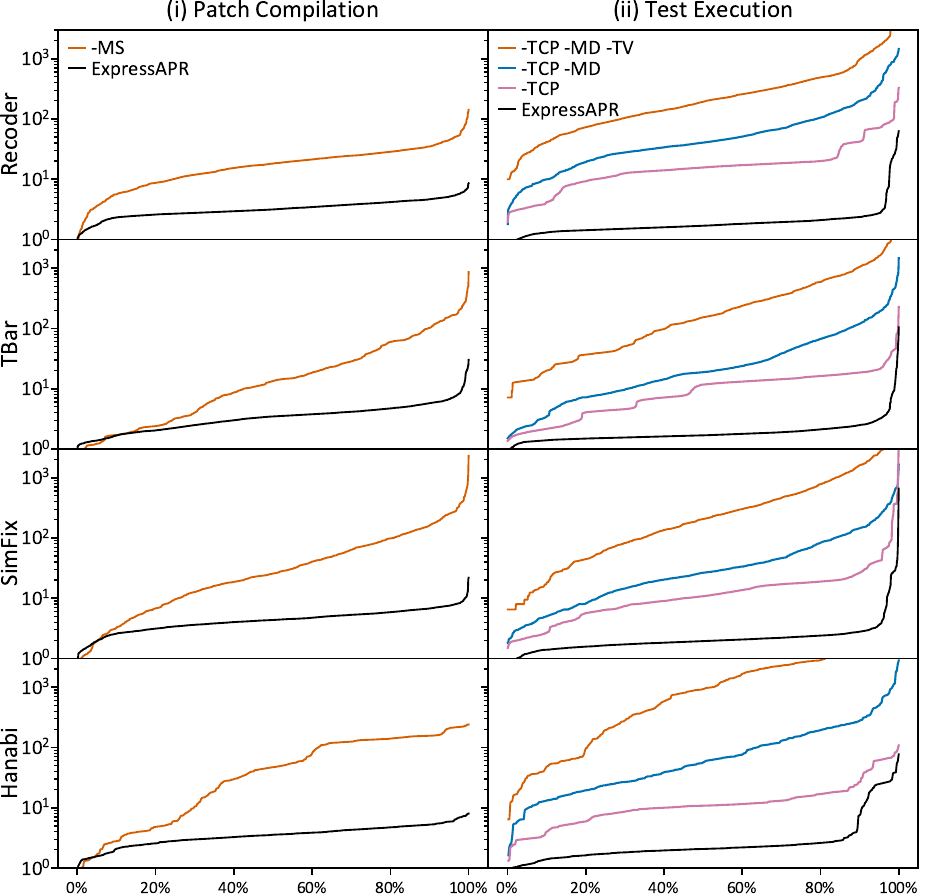}
        \vspace{-1.5em}
        \caption{fixable bugs}
    \end{subfigure}
    
    \vspace{.75em}
    
    \begin{subfigure}[b]{\columnwidth}
        \includegraphics[width=\columnwidth]{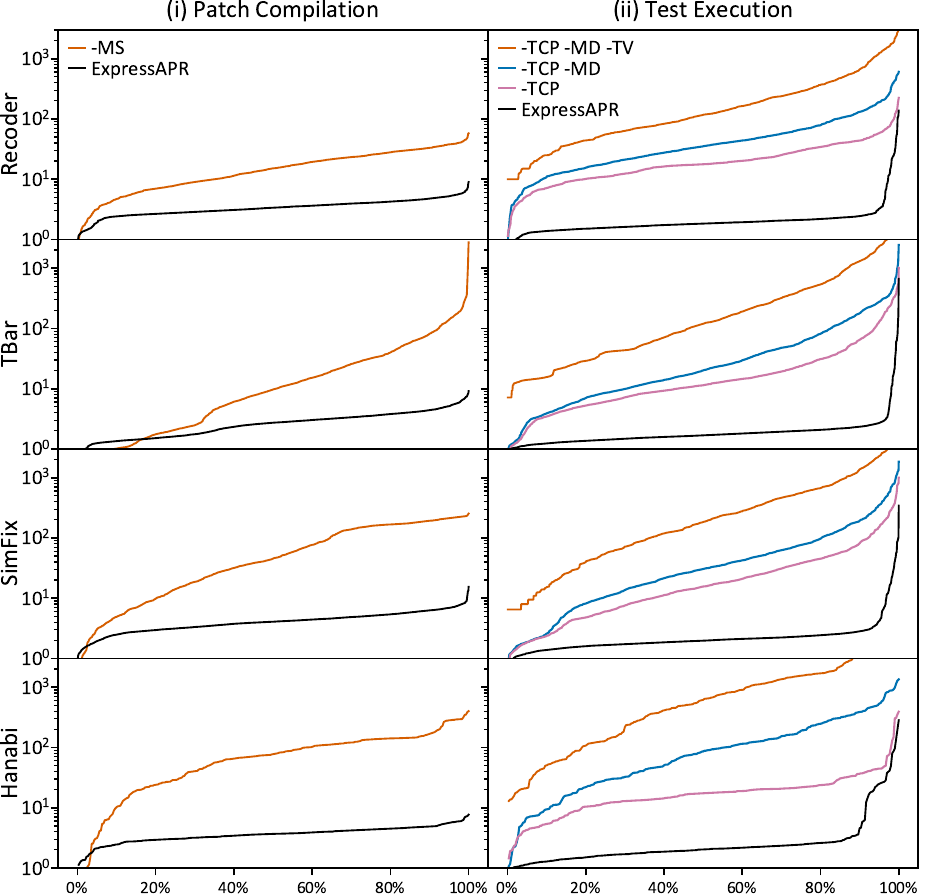}
        \vspace{-1.5em}
        \caption{random bugs}
    \end{subfigure}

    \caption{Effectiveness of each acceleration technique}
    \label{fig:result_speed_plot}
\end{figure}

\textbf{1. Acceleration techniques in ExpressAPR are effective on their own.}
\figref{fig:result_speed_plot} shows that every addition of technique contributes to a significant acceleration in its step. The average contribution of each technique is over 3x. Because the Y-Axis is logarithmic, the acceleration ratio can be directly read from the distance of two series.

\textbf{2. With all techniques used, patch compilation takes more time than test execution for most patch sets.} This can be confirmed by comparing the ExpressAPR series in (i) and (ii). This is because: 1) When TCP is used, most patches fail the first few unit tests, which costs little time. 2) The portion of patches that do not compile is considerable, precisely 54\%, 51\%, 76\%, and 52\% for the four APR tools, and test execution is skipped for them. It suggests that future work may spend more effort optimizing patch compilation.

\textbf{3. The effectiveness of Mutant Deduplication depends on the patch space.} We can see that MD performs best with Hanabi, by comparing the ``-TCP -MD'' and the ``-TCP'' series in different figures. This makes sense because Hanabi, as a decision-tree-based APR approach, naturally produces many patches changing Boolean conditions. These patches are more likely to be test-equivalent because Boolean conditions are either true or false.

\subsection{RQ3: Feasibility}

In this RQ, we count the number of cases where \ourtool fails to accelerate and classify them by their reasons. We also compare the validation result reported by \ourtool and the Plain baseline for detecting incorrect results.
The result is shown in \tabref{tab:result_feasibility}, which leads to the following findings:

\newcommand{\feasibilityBecause}[1]{ {\textit{#1}}}

\begin{table}[t]
    \centering\footnotesize
    \caption{The feasibility of \ourtool}
    \label{tab:result_feasibility}
    \begin{tabular}{@{\hspace{.25em}}l@{\hspace{1em}}r@{\hspace{1em}}r@{\hspace{.25em}}}
        \toprule
        \multirow{2}{*}{Category} & \multicolumn{2}{c}{\% of patches in ...} \\
        & fixable bugs & random bugs \\
        \midrule
        Correct result & 97.197\% & 98.782\% \\
        \midrule
        Validation failure \feasibilityBecause{(patch limitation)} & 1.882\% & 0.693\% \\
        \phantom{Validation failure} \feasibilityBecause{(test limitation)} & 0.886\% & 0.524\% \\
        \midrule
        Result misclassified \feasibilityBecause{(as plausible)} & 0.018\% & 0.000\% \\
        \phantom{Result misclassified} \feasibilityBecause{(as implausible)} & 0.017\% & 0.001\% \\
        \bottomrule
    \end{tabular}
\end{table}

\textbf{1. The acceleration feasibility of \ourtool is high.} In the fixable dataset and the random dataset, 97.197\% and 98.782\% of patches are correctly validated with acceleration. It indicates that the two implementation-level limitations do not affect the majority of patches.

\textbf{2. \ourtool has a negligible negative impact on the effectiveness of APR.} Most limitation violations are detected by \ourtool by reporting a validation failure. In our implementation, the plain validation approach is automatically used for these patches as a fallback, keeping the validation result correct. Only a very small portion of patches (0.035\% of the fixable dataset and 0.001\% of the random dataset) are misclassified because of undetected unstable tests. Among them, some are implausible patches misclassified as plausible, which can be ruled out by a post-check using the fallback approach on all plausible patches. So only plausible patches misclassified as implausible will have a negative impact on the effectiveness of APR. Since this portion is very small (0.017\% and 0.001\%), we believe the impact is negligible.

\subsection{RQ4: Generalizability}

\xyaadd{
When evaluating with Recoder under the IntroClassJava benchmark, \ourtool achieves an acceleration ratio of 41.5x. 99.77\% of patches can be accelerated. All accelerated patches have a correct patch validation result. Therefore:

\textbf{1. The acceleration of \ourtool generalizes to smaller programs.} The acceleration ratio under IntroClassJava (41.5x) is worse than Defects4J (108.9x for the random dataset), but is still very significant. The difference is possibly due to the fact that IntroClassJava is made of simple programs instead of large open-source projects. Therefore, test execution is less redundant with smaller codebases and fewer tests.

\textbf{2. The feasibility of \ourtool is improved for smaller programs.} Under IntroClassJava, \ourtool supports 99.77\% patches and achieves 100\% correctness, which is significantly better than the result under Defects4J (98.782\% supported for the random dataset). The improvement is intuitive, because simple programs are less likely to contain advanced syntax (breaking the patch limitation) or randomness (breaking the test limitation).
}

\subsection{Threats to Validity}

Threats to \textit{internal validity} might come from the possible faults when implementing \ourtool and performing the experiment. To avoid faults in our implementation, we have added assertions and sanity checks in the code. We have also manually inspected misclassified patches in RQ3 and did not find faults in the final implementation. To mitigate timing errors due to varying system load, we used the \code{cgroup} mechanism in Linux to assign one dedicated CPU core and enough RAM resources to each process.

Threats to \textit{external validity} lie in the representativeness of the benchmark. \xyaadd{To mitigate the threats, we experiment under both Defects4J (containing defects in open-source projects) and IntroClassJava (containing buggy student assignments), two widely used datasets for APR evaluation.} We collect candidate patches from four recent APR tools, covering multiple kinds of approaches. Therefore, our results have a high chance of representing the general use cases.


\section{Related Work}\label{section:related}

\subsection{Accelerating Patch Validation}

Existing approaches that accelerate patch validation can be categorized as \textit{general-purpose} or \textit{special-purpose}. General-purpose approaches~\cite{chen2021fast,ghanbari2020prf,guo2019speedup,mehne2018accelerating,weimer2013leveraging,qi2013efficient} take an arbitrary set of patches as input, while special-purpose approaches~\cite{wong2021varfix,ghanbari2019practical,hua2018practical,mechtaev2018test,chen2017contract,fast2010designing} are designed for and rely on a specific patch generation algorithm. \tabref{tab:related_work} summarizes acceleration techniques used in general-purpose approaches for easy comparison.
Below we compare our work against these existing approaches in three aspects.

\newcommand{\apsize}{\footnotesize}
\newcommand{\noname}[1]{\apsize\textit{{\kern -1pt}\citeauthor{#1}{\kern -1pt}}}

\begin{table}[t]
    \centering\footnotesize
    \caption{Acceleration techniques in general-purpose patch validators}
    \label{tab:related_work}
    \addtolength{\tabcolsep}{-5pt}
    \begin{tabular}{l|ccccccc}
    \toprule
Technique & \apsize \ourtool & \apsize UniAPR & \apsize PRF & DSU & \noname{mehne2018accelerating} & \apsize AE & \apsize FRTP \\
 & \textit{(ours)} & \cite{chen2021fast} & \cite{ghanbari2020prf} & \cite{guo2019speedup} & \cite{mehne2018accelerating} & \cite{weimer2013leveraging} & \cite{qi2013efficient} \\
\midrule
Mutant Schemata & \Checkmark &  &  &  & \#$^1$\hspace{-.5em} &  & \\
Mutant Dedup. & \Checkmark &  &  &  &  & & \\
Test Virtualization & \Checkmark & \Checkmark &  & \#$^2$\hspace{-.5em} &  &  & \\
Test Case Prio. & \Checkmark & \Checkmark & \Checkmark & \Checkmark & \Checkmark & \Checkmark & \Checkmark \\
Parallelization & \Checkmark & \#$^3$\hspace{-.5em} & \Checkmark & \#$^3$\hspace{-.5em} & \#$^3$\hspace{-.5em} & \Checkmark &\Checkmark \\
    \bottomrule
    \end{tabular}

    \vspace{.5em}
    {\raggedright\footnotesize
        \#$^1$: Does not discuss how to deal with uncompilable patches. \\
        \#$^2$: Reuses JVM without test virtualization, leading to incorrect results. \\
        \#$^3$: No built-in parallelization, but is parallelizable. \\
    }
    \addtolength{\tabcolsep}{+2pt}
\end{table}

1. Our empirical contribution is to understand the overall performance of integrating all five techniques and their relative accelerations on top of the other techniques. No existing approach has integrated all these techniques, so these empirical results were previously unknown. 

2. Our technical contribution is a set of novel techniques to overcome the challenges when adapting mutant schemata and mutant duplication to general-purpose patch validation. Three existing approaches have adapted mutant schemata within our knowledge. Two of them~\cite{chen2017contract,hua2018practical} are special-purpose: they are designed for a specific patch space that does not have the problem of uncompilable patches. Though the approach by \citet{mehne2018accelerating} is proposed as general-purpose, it does not discuss how to deal with uncompilable patches, which is one of our technical contributions. Four existing approaches have employed mutant deduplication within our knowledge. Three of them~\cite{weimer2013leveraging,hua2018practical,chen2017contract} statically detect the equivalence of patches and only reduce the patch validation time for fully equivalent patches. \citet{mechtaev2018test} uses the test-equivalence relationship to prune the patch space to avoid generating test-equivalent patches. However, their approach works for only a few special cases (interchangeable expressions and swappable statements) but cannot detect equivalence or test-equivalence in general (e.g., cannot detect \code{x+=2;} and \code{x++;x++;} are equivalent).

3. Some approaches~\cite{ghanbari2020prf,mehne2018accelerating,ghanbari2019practical,hua2018practical,fast2010designing} have adapted test case selection. We do not adapt this technique as it is subsumed by mutant deduplication, as discussed in \secref{section:survey_not_used}. 

\subsection{Mutation Testing}
\label{section:compare_mutation_testing}

There are multiple techniques for mutation testing acceleration, as surveyed in \secref{section:survey_used}. Our technical contribution is to adapt two classes of them, namely mutant schemata and mutant deduplication, to general-purpose patch validation. Below we compare our work against related mutation testing approaches.

1. Existing approaches with mutant schemata~\cite{untch1993mutation,just2011major,ma2005mujava} only allow pre-defined mutation operators against the program, which are designed not to cause compile errors. In comparison, we allow arbitrary changes to statements, so the space of mutation is significantly enlarged, and compile errors are handled by compilation isolation. \ourtool is more suitable for patch validation against mainstream APR tools, where the patch space is huge, and many patches cannot compile.

2. Among four existing approaches with mutant deduplication~\cite{baldwin1979heuristics,papadakis2015trivial,pan1994using,just2014efficient}, three approaches~\cite{baldwin1979heuristics,papadakis2015trivial,pan1994using} can detect only fully equivalent mutants, and are weaker than \ourtool, which detects test-equivalent mutants. The Major framework~\cite{just2014efficient} detects test-equivalence by interpreting mutants in a pre-pass. As discussed in \secref{section:approach_overview}, if a mutant's state transition deviates from the original states, it requires a static analysis process, which may be hard to implement precisely, or fail to detect some test-equivalent mutants if the analysis is imprecise. In comparison, our \napp approach does not need such analyses, so it is easier to implement and detect more test-equivalent mutants.

\subsection{The Effectiveness Aspect of APR}

\xyaadd{Many existing papers improve the effectiveness of APR by filtering or re-ranking candidate patches~\cite{ghanbari2022patch,xiong2018identifying,long2016automatic,ghanbari2020objsim}. While they may have a side-effect of improving efficiency (because plausible patches are ranked higher), we consider them orthogonal to our work. \ourtool performs only lossless acceleration and does not re-order patches. Therefore, \ourtool can be used together with such approaches to achieve a better performance.}

Also, readers may wonder about whether \ourtool can improve the effectiveness of APR (``Can we fix more bugs in the same time budget?''). While this is an interesting question, it is orthogonal to our work: we can already study this without \ourtool. \citet{vu2021please} has shown that given exponentially more time, the APR effectiveness only increases linearly or not at all. Therefore we do not expect that \ourtool can help APR tools repair significantly more bugs in its original time limit (generally several hours~\cite{noller2022trust}). That being said, \ourtool is useful for this kind of study, because the time required for experiments can be greatly reduced.

\section{Conclusion} \label{section:conclusion}


We surveyed mutation testing acceleration techniques and identified five classes of applicable techniques for general-purpose patch validation. We proposed two novel approaches, namely \napp and \eins, to overcome technical challenges when adapting two of the techniques for the first time.
Our large-scale empirical experiment has shown that patch validation can be dramatically accelerated and no longer be the time bottleneck. When acceleration techniques are systematically used, thousands of patches can be validated in minutes, satisfying the expectations of users.


\textbf{
The \ourtool artifact, including the source code with documentation, a command-line interface for APR users, a Docker image to reproduce the experiment, and raw experiment results, is available on GitHub~\cite{expressapr_replication}.
}

\renewcommand{\bibfont}{\footnotesize} 
\bibliographystyle{IEEEtranN}
\bibliography{IEEEabrv,bib/zotero,bib/base}

\newcommand{\biospace}{\vskip -2\baselineskip plus -1fil}

\biospace{}

\begin{IEEEbiography}[{\includegraphics[width=1in,height=1.25in,clip,keepaspectratio]{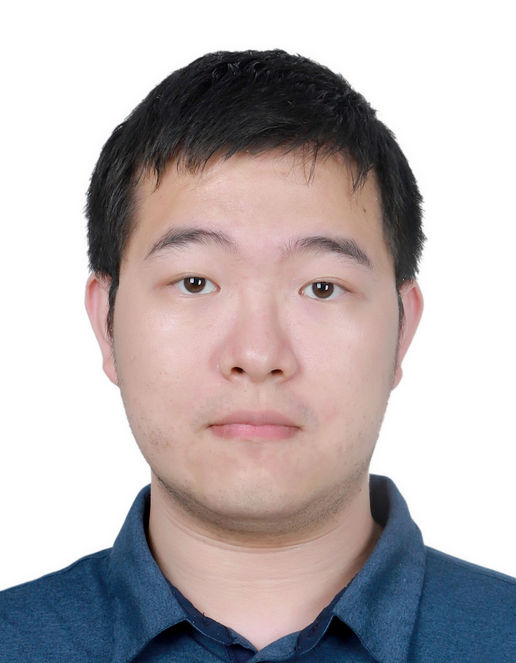}}]{Yuan-An Xiao}
is working toward the Ph.D. degree in computer science at Peking University.
He received his B.S. degree from the School of Electronics Engineering and Computer Science (EECS), Peking University. His research interests include program repair, software engineering, and cybersecurity.
\end{IEEEbiography}

\biospace{}

\begin{IEEEbiography}[{\includegraphics[width=1in,height=1.25in,clip,keepaspectratio]{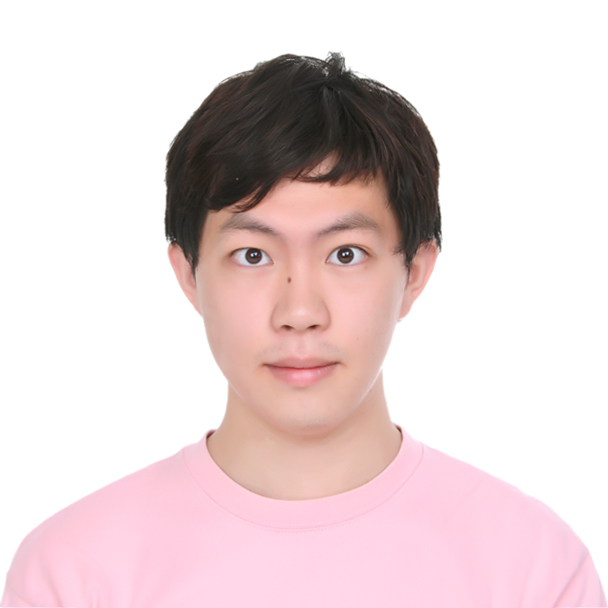}}]{Chenyang Yang}
is a Ph.D. student at Carnegie Mellon University.
He received his B.S. degree from the School of Electronics Engineering and Computer Science (EECS), Peking University. His research mainly focuses on software engineering and artificial intelligence.

\end{IEEEbiography}

\biospace{}

\begin{IEEEbiography}[{\includegraphics[width=1in,height=1.25in,clip,keepaspectratio]{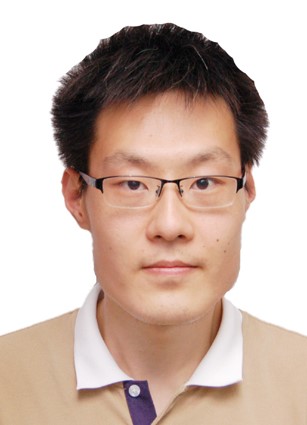}}]{Bo Wang} received the bachelor's degree from Central South University, Changsha, China, the master's degree from University of Science and Technology of China, Hefei, China, and the Ph.D. degree from Peking University, Beijing, China.
He is currently a Lecturer at the School of Computer and Information Technology, Beijing Jiaotong University, Beijing, China.
His current research interests include software testing and software analysis.
\end{IEEEbiography}

\biospace{}

\begin{IEEEbiography}[{\includegraphics[width=1in,height=1.25in,clip,keepaspectratio]{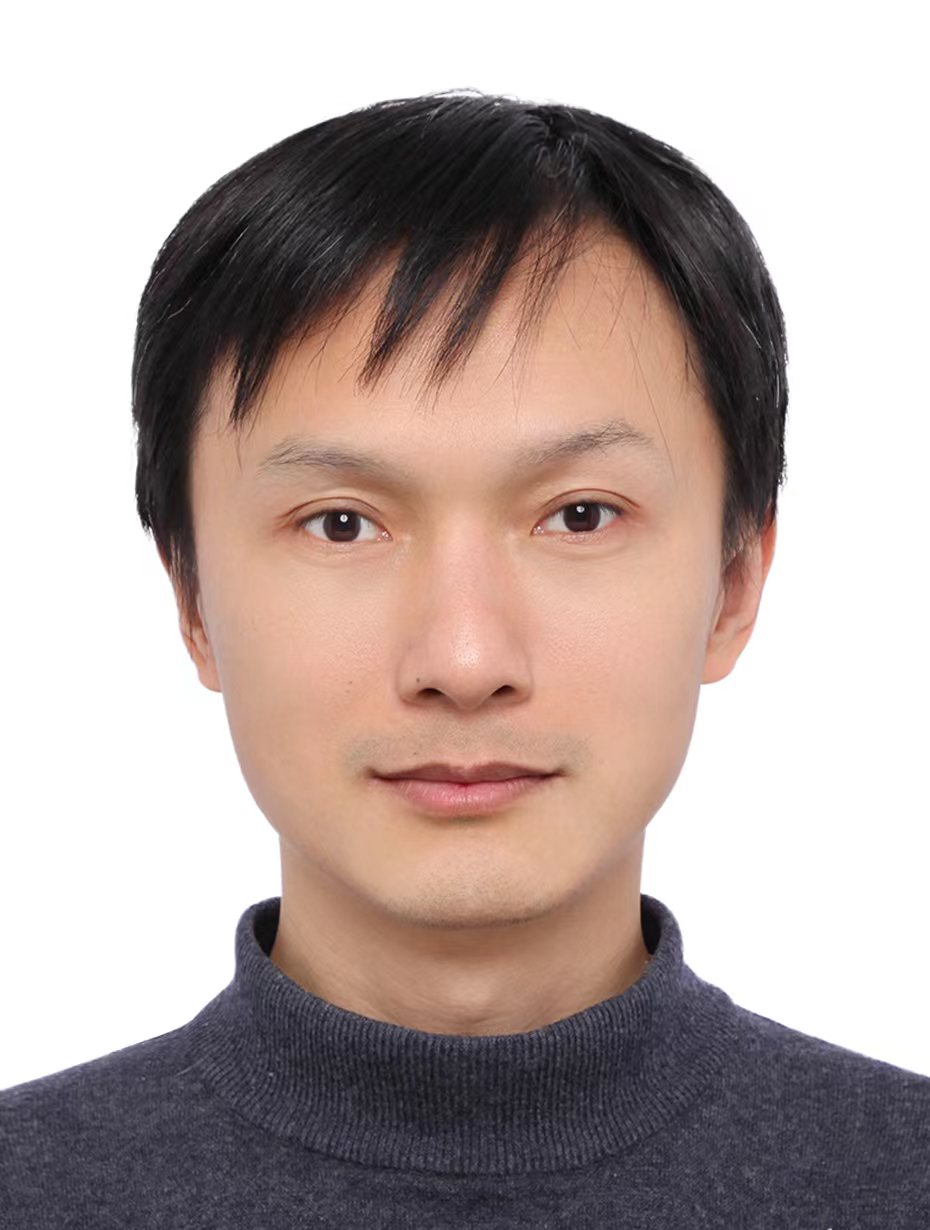}}]{Yingfei Xiong}
obtained his Ph.D. degree from the University of Tokyo in 2009, and worked as a postdoctoral researcher at University of Waterloo from 2009 to 2011. He joined Peking University in 2012 and is currently an associate professor. His research interest is programming languages and software engineering in general, and program synthesis, repair, and analysis in particular.
\end{IEEEbiography}

\end{document}